\documentclass[12pt]{article} 
\usepackage{amsfonts}
\usepackage{graphicx}

\makeatletter
\renewcommand{\@biblabel}[1]{${}^{#1}$}
\renewcommand{\@cite}[2]{{#1}}

\renewcommand{\@seccntformat}[1]{\csname the#1\endcsname. \hspace{0.5em}}
\makeatother

\newcommand{\CC}{{\mathbb C}}
\newcommand{\DDD}{{\cal D}}
\newcommand{\EE}{{\cal E}}
\newcommand{\NN}{{\mathbb N}}
\newcommand{\OO}{{\mathbb O}}
\newcommand{\bp}{{\bf p}}
\newcommand{\Q}{{\cal Q}}
\newcommand{\bq}{{\bf q}}
\newcommand{\RR}{{\mathbb R}}
\newcommand{\SM}{{\mathbb S}}
\newcommand{\br}{{\bf r}}
\newcommand{\UU}{{\mathbb U}}
\newcommand{\Uu}{{\cal U}}
\newcommand{\bx}{{\bf x}}
\newcommand{\by}{{\bf y}}
\newcommand{\bz}{{\bf z}}
\newcommand{\ba}{{\bf a}}
\newcommand{\ZZ}{{\mathbb Z}}
\renewcommand{\epsilon}{{\varepsilon}}
\renewcommand{\phi}{{\varphi}}
\newcommand{\halm}{{\vrule height7pt width4pt depth0pt}}
\newcommand{\Res}{{\rm Res}}
\newcommand{\dd}{\partial}
\newcommand{\bra}{\langle}
\newcommand{\ket}{\rangle}
\newcommand{\sh}{{\rm sh}}
\newcommand{\ch}{{\rm ch}}
\newcommand{\tgh}{{\rm th}}

\graphicspath{{prepare/}}

\begin{document}

\begin{center}\Large
Spectral properties of a short-range impurity in a quantum dot
\end{center}
\bigskip
 
\hskip 0.1\hsize
\begin{minipage}{0.8\hsize}
J.~Br\"uning,\footnote{Electronic mail: bruening@mathematik.hu-berlin.de}${}^)$
V.~Geyler,\footnote{On leave of absence from Laboratory of Mathematical Physics
Department of Mathematics, Mordovian State University, 430000 Saransk, Russia}${}^,$
\footnote{Electronic mail: geyler@mathematik.hu-berlin.de and geyler@mrsu.ru}${}^{)}$
and I.~Lobanov\footnote{Electronic mail: lobanov@mathematik.hu-berlin.de}${}^)$

{\it\small Institut f\"ur Mathematik, Humboldt-Universit\"at zu
Berlin, Unter den Linden 6, 10099 Berlin, Germany}
\bigskip\\
The spectral properties of the quantum mechanical system consisting of
a quantum dot with a short-range attractive impurity inside the dot are
investigated in the zero-range limit. The Green function of the
system is obtained in an explicit form. In the case of a
spherically symmetric quantum dot, the dependence of the
spectrum on the impurity position and the strength of the impurity
potential is analyzed in detail. 
It is proven that the confinement potential of the dot
can be recovered from the spectroscopy data.
The consequences of the hidden symmetry breaking by the impurity
are considered. The effect of the positional disorder is studied.
\end{minipage} 

\section{\label{sec:1}Introduction}
\indent
Quantum dots (i.e. nanostructures with charge carriers confinement
in all spatial directions) have an atom-like energy spectrum and
therefore make possible to fabricate quantum devices with energy
level spacing much greater than the temperature smearing $kT$ at
work temperature $T$ (see, e.g., Ref. \cite{BGL}). Moreover, dimension and shape of
a quantum dot affect considerably the most important
characteristics of the corresponding devices: relaxation and
recombination time, Auger recombination coefficient etc, thus a
possibility arises to control such characteristics in
manufacturing the devices.\textsuperscript{\cite{Har}, \cite{LUSKAB}, \cite{QD}}
Another way to control the properties of a quantum dot is
instilling an impurity into the dot. Therefore, the investigation
of spectral properties of a quantum dot with impurities as well as
the dependence of the spectrum on the geometric parameters of the
dot and physical characteristics of the impurity is an important
problem of nano- and mesoscopic physics (see, e.g., in Refs. \cite{FBP},
\cite{KGAS}, \cite{LGPW} and references therein). The case of a
hydrogen-like impurity is one of the most extensively studied up
to now; however, the spectral problem in this case has no exact
solution. On the other hand, short-range impurities can be
investigated in the framework of the point potential theory (also
called the zero-range potential theory). An important peculiarity
of the point potential method is that the spectral problem for a
point perturbed Hamiltonian is explicitly soluble as soon as the
Green function for the unperturbed operator is known in an
explicit form.\textsuperscript{\cite{AGHH},\cite{Pav}}

For modelling the geometric confinement of a quantum dot,
quadratic (in other words, parabolic) potentials are successfully
used\textsuperscript{\cite{Dav}} (see also examples of applications in Refs.
\cite{FBP}, \cite{KGAS}, \cite{LGPW}, \cite{BS} --
\cite{KGZ}). The reason is that the self-consistent
solution to the corresponding system of the Poisson and
Schr\"odinger equations leads to the confinement potential having
the form of a truncated parabolic potential.\textsuperscript{\cite{MF}}
Moreover, the Green function of the corresponding
Hamiltonian $\hat H^0$,
\begin{equation}
                              \label{HOS}
\hat H^0=-\frac{\hbar^2}{2\mu}\Delta+\frac{\mu\Omega^2}{2}r^2\,,
\end{equation}
can be explicitly calculated\textsuperscript{\cite{BVK}, \cite{GS}, \cite{KM}}
(here $\Omega$ is the frequency of the oscillator, $\mu$ denotes
over the paper the mass of the considered charged particle). This
makes possible to perform an exhaustive spectral analysis of the
perturbation of $\hat H^0$ by a point potential of arbitrary
position $\bq$ and strength $\alpha$ (we denote this perturbation
by $\hat H_\alpha(\bq)$) and to analyze the behavior of the
eigenvalues of $\hat H_\alpha(\bq)$ as functions of $\bq$ and
$\alpha$. This analysis is the main goal of the paper. Note that a
quite particular case of the point perturbation of $\hat H^0$ at
$\bq=0$ (without obtaining any explicit form for the Green
function) has been considered in Ref. \cite{FI96}. Point potential for
modelling an impurity in a spherically symmetric quantum dot has
been studied in the series of papers
using the Green function representation by means of the Laplace
transform of the propagator kernel, but this approach allows to
analyze (with numerical methods) the lowest impurity level 
only.\textsuperscript{\cite{KGAS}, \cite{KZE}, \cite{KZ}, \cite{KGZ}}

It should be noted that point perturbations of the one-dimensional
harmonic oscillators have been studied in detail earlier. This
study was started in Ref. \cite{APST}, where the spectral
properties of the point perturbed harmonic oscillator have been
considered in the context of the one-dimensional models for the
toponium physics and the Bose--Einstein condensation.\textsuperscript{\cite{PZ}} 
A strict mathematical justification of results from Ref. \cite{APST} was
done in Refs. \cite{FI94}, \cite{FI97}; see also in Ref. \cite{PWZ}.
Undoubtedly, our approach using the three-dimensional harmonic
oscillator is more adequate for the analyzing the spectral
properties of three-dimensional systems, in particular, the
toponium. It should be noted also that the one-dimensional
harmonic oscillator perturbed by a point potential with varying
position and strength has been investigated in Refs. \cite{GC97}, \cite{GC98}. 
A series of phenomena of low-dimensional condensed
matter physics can be analyzed by means of the Hamiltonian of the
perturbed oscillator: impurity in a one-dimensional quantum well,
one-dimensional channel in a two-dimensional heterostructure
subjected to a perpendicular uniform magnetic field etc., see the
bibliography in the cited papers for details. However, the
analysis given in Refs. \cite{GC97}, \cite{GC98} is based on the
properties of one-dimensional second-order differential operators
and is not extended to the three-dimensional case.

The paper is organized as follows. Preliminary results are
collected in Section \ref{sec:2}. In Section \ref{sec:3} we consider point
perturbations of the operator
\begin{equation}
                      \label{OP}
\hat H^0=-\frac{\hbar^2}{2\mu}\Delta+V\,
\end{equation}
with an infinitely growing potential $V$. It turns out that the
operator $\hat H_\alpha(\bq)$ can be defined and investigated for
the more generic case when $\hat H^0$ is defined by Equation
(\ref{OP}). In Section \ref{sec:4} some important properties of $\hat
H_\alpha(\bq)$ are established. In particular, a complete
description of the spectrum and eigenfunctions of $\hat
H_\alpha(\bq)$ is given in Theorem 1. As a consequence of this
theorem we get the falling of the considered particle on the
attractive center as the potential strength $\alpha$ tends to
$-\infty$; for a very particular case of the one-dimensional
harmonic oscillator perturbed at the potential minimum this
phenomenon was observed in Ref. \cite{APST}. In Section \ref{sec:5} we define at
fixed $\alpha$ a family of continuous functions such that the
values of these functions at the point $\bq$ form the complete
family of the eigenvalues of $\hat H_\alpha(\bq)$. Some elementary
properties of these functions are established in Theorem 2. The
main results of the paper are contained in Section \ref{sec:6}, where the
point perturbations of the Hamiltonian of the harmonic oscillator
are studied; the case of the isotropic harmonic oscillator
(\ref{HOS}) is considered in detail. These results are based on an
explicit form of the Green function for the operator (\ref{HOS}).
The detailed analysis of the dependence of the point levels on the
position $\bq$ and on the strength $\alpha$ is given in Theorem 3. In
particular, if $\bq\ne 0$, then the point levels never coincide
with the eigenvalues of the unperturbed operator $\hat H^0$.
Therefore we have here no accidental degeneracy of the levels,
which is a peculiarity of the one-dimensional model for the
toponium.\textsuperscript{\cite{APST}, \cite{GC97}} Hence, this degeneracy is an
artefact of the one-dimensional model. Another interesting results
is the asymptotic expression for the bound state of $\hat
H_\alpha(\bq)$ (Equations (\ref{as1}), (\ref{as2})). These
equations  show that at least for the isotropic harmonic oscillator
its potential (i.e., the frequency $\Omega$) can be recovered from
the dependence of the ground state of the point perturbation on
the support of the perturbation. Moreover, we argue that the form
of the parabolic potential $V$ may be recovered from the behavior
of the excited energy for the ground state. Our conjecture is that
this property is true for a more general form of the potential
$V$. In this connection it is of interest to note that the study
of the excited energy is one of the main problems of the quantum
dot physics.\textsuperscript{\cite{KGAS}} The methods of Section \ref{sec:6} allow us to analyze
rigorously the phenomenon of so-called "positional disorder" in
quantum dots (including nonisotropic ones). The relation of the
degeneracy properties of the eigenvalues of $\hat H_\alpha(\bq)$
at $\bq=0$ to the symmetry properties of the unperturbed operator
$\hat H^0$ in the phase space is briefly discussed in the
conclusion of Section \ref{sec:6}. In particular, the appearance of states with
nonzero dipole momentum is noted.

\section{\label{sec:2}Preliminaries}
\indent
Here we present for the convenience of readers some
basic properties of point perturbations of Schr\"odinger operators
in $L^2(\RR^3)$ (see, e.g., Refs. \cite{AGHH}, \cite{BG}, \cite{CFKS},
\cite{GMC1}, \cite{Sim} for details). We will consider only
Schr\"odinger operators $\hat H^0$ of the form (\ref{OP}), where
the potential $V$ is subordinated to the conditions

\begin{itemize}

\item[(P1)] $V\in L^p_{\rm loc}(\RR^3)$ {\it for some $p>3$};

\item[(P2)] $V_-=\min(V,0)\in L^2(\RR^3)+L^\infty(\RR^3)$.
\end{itemize}

Conditions (P1), (P2) are weaker than commonly used in
applications conditions $V\in L^\infty_{\rm loc}(\RR^3)$ and $V\ge
c$ with $c\in\RR$ but making use of (P1), (P2) requires no change
in proving of main results below. It is well known that under
these conditions $\hat H^0$ is semibounded from below and
essentially self-adjoint on $C_0^\infty(\RR^3)$ (see in Ref. \cite{RS2}
Theorem X.28). Further we put, as a rule, $\hbar=1$, $\mu=1/2$ and
denote the obtained operator $-\Delta+V$ by $H^0$. For the domain
$\DDD(H^0)$ of $H^0$ we have $C_0^{\infty}(\RR^3)\subset
\DDD(H^0)\subset C(\RR^3)$. This inclusion implies that the Green
function $G^0(\bx,\by;\zeta)$ for $H^0$ (the integral kernel of
the resolvent $R(\zeta)= (H^0-\zeta)^{-1}$) is a Carleman operator,
this means that
\begin{equation}
                   \label{Car}
\int\limits_{\RR^3}\,|G^0(\bx,\by;\zeta)|^2\,d\by\,<\,+\infty\quad\,{\rm
for}\,\,\,{\rm  a.e.}\,\,\,\bx\in\RR^3\,.
\end{equation}
Moreover, according to Theorem B.7.2 from Ref. \cite{Sim}, for every
fixed $\zeta$, $\zeta\in\CC\setminus{\rm spec}\,(H^0)$, the function $G^0$
obeys the following properties:

\begin{itemize}

\item[(G1)] {\it For every
$\zeta\in{\rm spec}(H^0)$ the function $G^0(\bx,\by;\zeta)$ is
continuous in the domain
$\{(\bx,\by)\in\RR^3\times\RR^3\,:\bx\ne\by\}$};

\item[(G2)] $|G^0(\bx,\by;\zeta)|\le c_2(\zeta)|\bx-\by|^{-1}$;

\item[(G3)] {\it if $|\bx-\by|\ge d>0$, then $|G^0(\bx,\by;\zeta)|\le
c_3(d,\delta,\zeta)
\exp(-\delta|\bx-\by|)$ for some $\delta>0$. Moreover, if ${\rm
Re}\,\zeta<\Sigma\equiv \inf{\rm spec}(H^0)$, then arbitrary
$\delta$ with $\delta^2/2< \Sigma-{\rm Re}\,\zeta$ is suitable for
this estimate.}
\end{itemize}
\noindent From (G1) we get, in particular, that (\ref{Car}) is
valid for {\it every} $\bx\in\RR^3$.

The crucial role in the point potential theory is played by the
regularized Green function
\begin{equation}
                                     \label{GS}
G^0_{\rm reg}(\bx,\by;\zeta)=G^0(\bx,\by;\zeta)-
\frac{1}{4\pi}\frac{1}{|\bx-\by|}\,.
\end{equation}
In the particular cases, e.g. if $V\in C^\infty(\RR^3)$, it is
known that at fixed $\zeta$ this function has a continuous
extension on the whole space $\RR^3\times\RR^3$  (see, e.g.,
Ref. \cite{Mir} or Theorem III.5.1 in Ref. \cite{Ber}). We need this property
in the general situation and prove it under conditions (P1), (P2).

It is sufficient to prove that $G^0_{\rm reg}(\bx,\by;\zeta)$ is
continuous with respect to $(\bx,\by)$ for some $\zeta=E_0<0$.
Indeed, then for every $\zeta\in\CC\setminus{\rm spec}\,(H^0)$
$$
G^0(\bx,\by;\zeta)=\int\limits_{E_0}^\zeta\,\frac{\dd}{\dd\lambda}
G^0(\bx,\by;\lambda)\,d\lambda+G^0(\bx,\by;E_0)\,,
$$
where the path of integration lies in the resolvent set
$\CC\setminus{\rm spec}\,(H^0)$. The function\linebreak
$(\dd G^0/\dd\lambda)(\bx,\by;\lambda)$ is jointly
continuous with respect to $(\bx,\by)$ since it coincides with the
integral kernel of $(H^0-\lambda)^{-2}$ and this kernel is
continuous according to  Theorem B.7.1 from Ref. \cite{Sim}.

It is easy to see that $V$ can be represented in the form
$V=V_1+W$, where $V_1\in C^\infty(\RR^3)$ and obeys the property
(P2) and $W\in L^p(\RR^3)\cap L^1(\RR^3)$.
Denote $H^1=-\Delta+V_1$, $\Sigma_1=\inf{\rm spec}\,(H^1)$ and by
$G^1$ the Green function of $H^1$. Fix $E_0$,
$E_0<\min(\Sigma,\Sigma^1)$, and introduce the function
$F(\bx,\by,\bz)=G^0(\bx,\bz;E_0)W(\bz)G^1(\bz,\by;E_0)$. Using
properties (G2), (G3) and the estimate
\begin{equation}
                  \label{In}
\int\limits_{|\by-\ba|\le r}\frac{d\by}{|\bx-\by|^\nu}\le \tilde
c_\nu r^{3-\nu}\,,
\end{equation}
where $0<\nu<3$, $r>0$, $\ba,\,\bx\in\RR^3$, it is easy to prove
that $F(\bx,\by,\cdot)\in L^1(\RR^3)$ for all $\bx,\by\in\RR^3$.
In virtue of the Lippmann--Schwinger relation
$$
G^0(\bx,\by;E_0)=G^1(\bx,\by;E_0)+\int\limits_{\RR^3}\,
G^0(\bx,\bz;E_0)W(\bz)G^1(\bz,\by;E_0)\,d\bz
$$
and the continuity of the regularized Green function for $H^1$, it
remains to prove that the function
$$
I(\bx,\by)=\int\limits_{\RR^3}F(\bx,\by,\bz)\,d\bz
$$
is continuous on $\RR^3\times\RR^3$. Moreover, (G1) shows that it
remains to prove the continuity of $I$ at points of the form
$(\bx_0,\bx_0)$. To do this fix $\epsilon>0$ and find $\eta>0$
such that the relations $|\bx-\bx_0|<\eta$, $|\by-\bx_0|<\eta$
imply $|I(\bx,\by) -I(\bx_0,\bx_0)|\le \epsilon$. Introduce  the
sets $B_1(\eta)=\{\bz:\,|\bz-\bx_0|<\eta\}$,
$B_2(\eta)=\RR^3\setminus B_1(\eta)$, and for a measurable set
$B\subset \RR^3$ denote $I_B(\bx,\by)=
\int_B\,F(\bx,\by,\bz)\,d\bz$. Then
$$
|I(\bx,\by) -I(\bx_0,\by_0)|\le |I_{B_1(\eta)}(\bx,\by)|+
|I_{B_1(\eta)}(\bx_0,\by_0)| +|I_{B_2(\eta)}(\bx,\by)
-I_{B_2(\eta)}(\bx_0,\by_0)|\,.
$$
If $\bx,\by,\bz\in B_1(\eta)$, then by (G2)
$$
|F(\bx,\by,\bz)|\le f(\bz)\,|\bx-\by|^{-1}|\bz-\by|^{-1}\,
$$
where $f\in L^p$, therefore relation (\ref{In}) and the
Cauchy--Schwartz inequality lead to the estimate
$|I_{B_1(\eta)}(\bx,\by)|+ |I_{B_1(\eta)}(\bx_0,\by_0)|\le {\rm
const}\,\eta$. On the other hand, if $\bx,\by\in B_1(\eta/2)$,
$\bz\in B_2(\eta)$, then we have from (G3):
$|F(\bx,\by,\bz)|\le g(\bz)\,\exp(-\delta|\bz|)$, where $\delta>0$
and $g\in L^p$. Thus by (G1) and the Lebesgue majorization
theorem, $I_{B_2(\eta)}(\bx,\by)$ is a continuous function on
$B_1(\eta/2) \times B_1(\eta/2)$, and the proof of continuity of
$G^0_{\rm reg}$ is completed.

\medskip

Let $\bq\in\RR^3$, then the restriction of $H^0$ to the domain
$\{f\in\DDD(H^0)\,:\,f(\bq)=0\}$ is a closed symmetric operator
$S$ with the deficiency indices $(1,1)$. By definition, the {\it
point perturbation of $H^0$, supported on $\bq$} is a self-adjoint
extension of $S$ different from $H^0$. All the point perturbations
of $H^0$ supported on a given $\bq\in\RR^3$ form a one-parameter
family $H_{\alpha}(\bq)$, $\alpha\in\RR$, of self-adjoin operators
such that the Green function $G_{\alpha}$ of $H_{\alpha}(\bq)$ is
given by the formula
\begin{equation}
                        \label{Kr}
G_{\alpha}(\bx,\by;\zeta)= G^0(\bx,\by;\zeta)-
\left[Q(\zeta;\bq)-\alpha\right]^{-1}G^0(\bx,\bq;\zeta)G^0(\bq,\by;\zeta)\,,
\end{equation}
which is a consequence of the Krein resolvent formula. Here
$Q(\zeta;\bq)=G^0_{\rm reg}(\bq,\bq;\zeta)$ is the so-called Krein
$\Q$-function. The operator $H^0$ corresponds formally to
$\alpha=\infty$; moreover, $H^0$ is the Friedrichs extension of
$S$.

The extension parameter $\alpha$ has an important physical
meaning, namely, $H_\alpha$ can be treated as the Hamiltonian
$H^0$ perturbed by a zero-range potential, in this case $\alpha$
is the strength of this potential.\textsuperscript{\cite{AGHH}, \cite{BZP},
\cite{DO}} In place of the strength $\alpha$, it is more
convenient to use for applications so-called "scattering length"
$\ell_s$, $\ell_s=1/(4\pi\alpha)$ 
(see in Refs. \cite{AGHH}, \cite{BZP}, \cite{DO} again). More precisely,
$$
\ell_s=\frac{\mu}{2\pi\hbar^2\alpha}\,,
$$
and we see that $\ell_s$ has actually the dimension of the length.

Note that according to the general results of the
Krein self-adjoint extension theory, the function $\zeta\mapsto
Q(\zeta;\bq)$ is analytic in the domain $\CC\setminus{\rm
spec}(H^0)$ for each $\bq\in\RR^3$
and $\dd Q(E;\bq)/\dd E>0$
if $E\in\RR\setminus{\rm spec}(H^0)$.\textsuperscript{\cite{KL}}
Remark that $Q(\zeta;\bq)$ can be continuously extended to some
points of ${\rm spec}(H^0)$. Further we assume that $Q(\zeta;\bq)$
is continuously extended to all regular points.

It is easy to prove that for every $\bq\in\RR^3$ the mapping
$\zeta\mapsto G^0(\cdot,\bq;\zeta)$ is an analytic function from
the domain $\CC\setminus{\rm spec}\,(H^0)$ to the Hilbert space
$L^2(\RR^3)$. Denote $G^0(\cdot,\bq;\zeta)$ by $g_{\bq}(\zeta)$,
then we can rewrite (\ref{Kr}) in an operator form
\begin{equation}
                        \label{KrOp}
R_{\alpha}(\zeta)=
R^0(\zeta)-
\left[Q(\zeta;\bq)-\alpha\right]^{-1}|g_{\bq}(\zeta)\ket\bra g_{\bq}(\zeta)|\,,
\end{equation}
where $R_{\alpha}(\zeta)=(H_{\alpha}-\zeta)^{-1}$ and
$R^0(\zeta)=(H^0-\zeta)^{-1}$.

Note, that $g_{\bq}(\zeta)$ is a nonzero function for every
$\bq\in\RR^3$ and $\zeta\in\CC\setminus{\rm spec}\,(H^0)$. Indeed,
otherwise we have $\varphi(\bq)=0$ for every $\varphi\in
\DDD(H^0)$ that contradicts the inclusion
$C_0^{\infty}(\RR^3)\subset \DDD(H^0)$.

In conclusion we mention a possibility to approximate the
zero-range perturbation by potentials with decreasing support. For
$V=0$ the corresponding procedure is described in Ref. \cite{AGHH}
(Theorem 1.2.5). We sketch here the proof for $H^0$ with potential
$V$ having properties (P1), (P2).

Let $W\in L^2_{\rm comp}(\RR^3)$, in particular, $W$ is a Rollnik
function (see in Ref. \cite{RS2}, Sec. X.2). Denote $v=|W|^{1/2}$, $u=v\,{\rm
sign}(V)$, and let $\lambda(\epsilon)$ be a real-analytic function
in a neighborhood of zero such that $\lambda(0)=1$. For
$\epsilon>0$ consider the operator $H^\epsilon\equiv
H^\epsilon(\bq)=H^0+\epsilon^{-2}\lambda(\epsilon)W(\epsilon^{-1}(\bx-\bq))$.
Then the resolvent $R^\epsilon(\zeta)=(H^\epsilon-\zeta)^{-1}$
$(\epsilon>0)$ has the form
$$
R^\epsilon(\zeta)=R^0(\zeta)-\epsilon\lambda(\epsilon)A^\epsilon[1+B^\epsilon
]^{-1}C^\epsilon\,,
$$
where $A^\epsilon$, $B^\epsilon$, $C^\epsilon$ are integral
operators with the kernels $A^\epsilon(\bx,\by;\zeta)= G^0(\bx,
\epsilon\by+\bq;\zeta)v(\by)$, $C^\epsilon(\bx,\by;\zeta)=
G^0(\epsilon\bx+\bq,\by;\zeta)u(\bx)$,
$B^\epsilon(\bx,\by;\zeta)=\epsilon\lambda(\epsilon)
G^0(\epsilon\bx+\bq,\epsilon\by+\bq;\zeta)u(\bx)v(\by)$. Define
$A^0$ and $C^0$ putting  $\epsilon=0$ in the formulas above, and
define $B^0$ by the integral kernel
$B^0(\bx,\by)=(4\pi|\by-\bx|)^{-1}u(\bx)v(\by)$. All the operators
$A^\epsilon$, $B^\epsilon$ and $C^\epsilon$  ($\epsilon\ge 0$)
belong to the Hilbert--Schmidt class and $A^\epsilon\to A^0$,
$B^\epsilon\to B^0$, $C^\epsilon\to C^0$ with respect to the
Hilbert--Schmidt norm as $\epsilon\to+0$. Moreover, 
using (\ref{GS}) we can prove that with respect to this norm
$$
B^\epsilon=B^0+\epsilon(\lambda'(0)B^0+Q(\zeta;\bq)|u\ket\bra
v|)+o(\epsilon)\,.
$$
Hence, the arguments using for the proof of Theorem 1.2.5 from
Ref. \cite{AGHH} give the following result.

\bigskip

\noindent{\bf Theorem A.}

\begin{itemize}

\item[(1)] {\it Let $\bra v|\phi\ket=0$ for all $L^2$-solutions $\phi$ of the equation $B^0\phi=-\phi$
$($in particular, let $-1$ be not an eigenvalue of $B_0$$)$. Then
$H^\epsilon(\bq) \to H^0$ in the norm-resolvent sense as
$\epsilon\to+0$.}

\item[(2)] {\it Let $-1$ be a simple eigenvalue of $B^0$ and $\phi$ be a corresponding
eigenfunction normalized by the condition $\bra\tilde
\phi|\phi\ket=-1$, where $\tilde \phi=\phi\,{\rm sign}(V)$. If
$\bra v|\phi\ket\ne0$, then
$\displaystyle\lim\limits_{\epsilon\to+0} H^\epsilon(\bq)=
H_\alpha(\bq)$ in the norm-resolvent sense, where
$\alpha=-\lambda'(0)|\bra v|\phi\ket|^{-2}$}.

\item[(3)] {\it Let $-1$ be a multiple eigenvalue of $B^0$ with eigenfunctions $\phi_1,\ldots,\phi_n$
normalized by the conditions $\bra\tilde
\phi_j|\phi_k\ket=-\delta_{jk}$ $(\tilde \phi_j=\phi_j\,{\rm
sign}(V))$. If $\bra v|\phi_j\ket\ne0$ for some $j$ and
$\lambda'(0)\ne0$, then $\displaystyle\lim\limits_{\epsilon\to+0}
H^\epsilon(\bq)= H_\alpha(\bq)$ in the norm-resolvent sense, where
$$\alpha=-\lambda'(0)\displaystyle\left[\sum_{j=1}^n|\bra
v|\phi_j\ket|^{2}\right]^{-1}.\halm$$} 

\end{itemize}

\section{\label{sec:3}Point perturbation in the case of unbounded potential $V$}

Starting with this section we suppose additionally that

\begin{itemize}

\item[(P3)] $ \lim\limits_{|\br|\to\infty} V(\br)=+\infty$.

\end{itemize}

\noindent In this case $R^0(\zeta)$ is a compact operator for all
$\zeta\in\CC\setminus{\rm spec}(H^0)$ (the Strichartz theorem; see,
e.g., in Ref. \cite{RS4}, Theorem XIII.69). Therefore ${\rm spec}(H^0)$
consists of an unbounded sequence
$\lambda_0<\lambda_1<\ldots<\lambda_n<\ldots$ of eigenvalues with
finite multiplicity $k_n$. Consequently, $Q(\zeta;\bq)$ is a
meromorphic function of $\zeta$. We are going to find the poles of
this function.

Denote by $L_n$ the eigenspace
associated with $\lambda_n$, and choose in $L_n$ an
orthonormal basis $F_{n,k}(\br)$, $k=1,\ldots,k_n$.
For every $\bq\in\RR^3$ we denote
$$
\sigma(\bq)=\{\lambda_n\in {\rm spec}(H^0): \exists f\in L_n\quad
{\rm s.t.}\quad f(\bq)\ne 0\}
$$

\bigskip

\noindent{\bf Lemma 1.} {\it The set of all poles of the function
$\zeta\mapsto Q(\zeta;\bq)$ coincides with $\sigma(\bq)$}.

\medskip

\noindent{\bf Proof.} Since $(\dd G^0/\dd\zeta)(\bx,\by;\zeta)$ is the integral kernel for the operator
$(H^0-\zeta)^{-2}$, we have according to the Mercer theorem
$$
\frac{\dd}{\dd
\zeta}G^0(\bx,\by;\zeta)=\sum\limits_{n=0}^{\infty}\sum\limits_{k=1}^{k_n}\,
(\lambda_n-\zeta)^{-2}F_{n,k}(\bx)\overline{F_{n,k}(\by)}\,,
$$
where the series converges locally uniformly on
$\RR^3\times\RR^3\times(\CC\setminus{\rm spec}(H^0))$. Therefore,
\begin{equation}
            \label{QS}
\frac{\dd}{\dd
\zeta}Q(\zeta;\bq)=\sum\limits_{n=0}^{\infty}\sum\limits_{k=1}^{k_n}\,
(\lambda_n-\zeta)^{-2}|F_{n,k}(\bq)|^2\,,
\end{equation}
and the series converges locally uniformly on
$(\CC\setminus{\rm spec}(H^0))\times\RR^3$.
The lemma follows from (\ref{QS}) immediately. \halm

\bigskip

\noindent{\bf Lemma 2.} {\it For each $\bq\in\RR^3$ the set
$\sigma(\bq)$ is infinite. If $V$ is bounded from below, then
$\lambda_0\in\sigma(\bq)$.}

\medskip

\noindent{\bf Proof.} Consider the space of continuous functions
$C(\RR^3)$ with the topology of compact convergence. Due to the
closed graph theorem and the relation $\DDD(H^0)\subset C(\RR^3)$,
the operator $R^0(-1):\,L^2(\RR^3)\rightarrow C(\RR^3)$ is
continuous. Therefore for every $f\in\DDD(H^0)$ the Fourier
expansion for $f$ with respect to the basis $(F_{n,k})_{n,k}$
converges locally uniformly. Assume that the set $\sigma(\bq)$ is
finite; let $N=\max\{n:\,\lambda_n\in \sigma(\bq)\}$ and $P$ be
the orthogonal projection of $L^2(\RR^3)$ on the subspace $M=
L_0+\ldots+L_N$. Then for every $\phi\in\DDD(H^0)$ the conditions
$\phi(\bq)=0$ and $(P\phi)(\bq)=0$ are equivalent. Since $M$ is
finite dimensional, there is $h\in M$ such that for every $\phi\in
M$ the conditions $\phi(\bq)=0$ and $\bra h\,|\,\phi\ket=0$ are
also equivalent. Using the inclusion $C_0^{\infty}(\RR^3)\subset
\DDD(H^0)$ we see that there is a function $h\in L^2(\RR^3)$ such
that for every $\phi\in C_0^{\infty}(\RR^3)$ the conditions
$\phi(\bq)=0$ and $\bra h\,|\,\phi\ket=0$ are equivalent.
Obviously, this is impossible, hence $\sigma(\bq)$ is infinite. If
$V$ is bounded from below, then by Theorem XIII.48 from Ref. \cite{RS4}
the eigenfunctions of $H^0$ corresponding to the ground state
$\lambda_0$ have no zeros therefore $\lambda_0\in\sigma(\bq)$.
\halm

\bigskip

Another property of the function $\zeta\mapsto Q(\zeta;\bq)$ we
need further follows.

\medskip

\noindent{\bf Lemma 3.} {\it The function $Q(\zeta;\bq)$ tends to
$-\infty$ as $\zeta\rightarrow-\infty$, $\zeta\in\RR$.}

\medskip

\noindent{\bf Proof.} Since $H^0$ is the Friedrichs extension of
$S$, the statement follows from Proposition 4 of Ref. \cite{DM}. \halm

\section{\label{sec:4}Spectral properties of $H_\alpha$ at fixed position of the
point perturbation}

Here we describe the spectrum of $H_\alpha(\bq)$ for a
fixed $\bq\in\RR^3$. Further, if it does not lead to a
misunderstanding, we omit $\bq$ from the notations.

Since $H_\alpha$ is a rank one perturbation of $H^0$, the spectrum
of $H_\alpha$ is discrete. Moreover, an eigenvalue $\lambda_n$ of
$H^0$ of the multiplicity $k_n$ is an eigenvalue of $H_\alpha$ of
the multiplicity $k_n-1$, $k_n$ or $k_n+1$ (if $k_n=1$, the first
case means, of course, that $\lambda_n$ does not belong to ${\rm
spec}(H_\alpha)$). For $\lambda\notin {\rm spec}(H^0)$ we see from
(\ref{KrOp}) that $\lambda$ is an eigenvalue of $H_\alpha$ if and
only if $\zeta=\lambda$ is a solution to the equation

\begin{equation}
                           \label{SE}
Q(\zeta;\bq)-\alpha=0.
\end{equation}

Denote by $(\epsilon_n)_{n\in\NN}=(\epsilon_n(\bq))_{n\in\NN}$ the
strictly increasing sequence of all the poles of $Q(\zeta;\bq)$.
Since $(\dd Q/\dd E)(E;\bq)>0$ for
$E\in\RR\setminus{\rm spec}(H^0)$, the equation (\ref{SE})
has exactly one solution on each interval
$(-\infty,\epsilon_0),(\epsilon_0,\epsilon_1),\ldots$ Denote such
solutions, which do not belong to ${\rm spec}(H^0)$, by
$\EE_0,\EE_1,\ldots$, where $\EE_0<\EE_1<\cdots$. The following
theorem completely describes the eigenvalues and the
eigenfunctions of $H_\alpha(\bq)$.

\bigskip

\noindent{\bf Theorem 1.} {\it Let $\bq\in\RR^3$ be fixed. The
spectrum of $H_\alpha=H_\alpha(\bq)$ is discrete and consists of
four nonintersecting parts $\sigma_1, \sigma_2, \sigma_3,
\sigma_4$ described as follows.

\begin{enumerate}

\item[$(1)$] $\sigma_1$ is the set of all solutions $\EE_n$ to the equation $($\ref{SE}$)$,
which do not belong to ${\rm spec}(H^0)$.
The multiplicity of $\EE_n$ in the spectrum of $H^0$ is equal to $1$.

\item[$(2)$] $\sigma_2$ is
the set of all $\lambda_n\in\sigma(\bq)$ that are
multiple eigenvalues of $H^0$.
The multiplicity of the eigenvalue $\lambda_n\in\sigma_2$
in the spectrum of $H_\alpha$ is equal to $k_n-1$.

\item[$(3)$] $\sigma_3$ consists of all $\lambda_n$,
$\lambda_n\in{\rm spec}\,(H^0)\setminus\sigma(\bq)$, that are not
solutions of $($\ref{SE}$)$. The multiplicity of the eigenvalue
$\lambda_n$ in ${\rm spec}(H_\alpha)$ is equal to $k_n$.

\item[$(4)$] $\sigma_4$ consists of all $\lambda_n$,
$\lambda_n\in{\rm spec}\,(H^0)\setminus\sigma(\bq)$, such that
$\lambda_n$ is a solution of $($\ref{SE}$)$. The multiplicity of
the eigenvalue $\lambda_n$ in ${\rm spec}(H_\alpha)$ is equal to
$k_n+1$.

\end{enumerate}

The corresponding eigensubspaces are described as follows.

\begin{enumerate}

\item[$(1)$] The subspace spanned by the normalized eigenfunction

$$
\Phi_n=\left[\frac{\partial Q}{\partial
\zeta}(\EE_n;\bq)\right]^{-\frac{1}{2}} g_{\bq}(\EE_n)\,.
$$

\item[$(2)$] The orthogonal complement
in $L_n$ of the function
$$
\Psi_n(\bx)=\sum_{k=1}^{k_n}\overline{F_{n,k}(\bq)}F_{n,k}(\bx)\,,
$$
or, equivalently, the subspace of $L_n$ of the form $\{f\in
L_n:\,f(\bq)=0\}$.

\item[$(3)$] The subspace $L_n$.

\item[$(4)$] The direct sum of $L_n$ and the space spanned by
the function $g_{\bq}(\lambda_n)$, which is orthogonal to $L_n$.

\end{enumerate}
}

\medskip

\noindent{\bf Proof.}\,\,
The proof is based on direct calculations with the help of
following statements:

\medskip

\noindent(A) {\it The orthoprojector $P(E_0)$ on the eigenspace
of a self-adjoint operator $T$ corresponding to an isolated eigenvalue $E_0$
has the form}
$$
P(E_0)=-\Res[(T-\zeta)^{-1};\zeta=E_0].
$$

\medskip

\noindent(B) {\it Suppose $P_1, P_2$ and $P_1+cP_2$, where
$c\in\CC$, are orthoprojectors in a Hilbert space and $P_2\ne 0$,
then $c$ equals $0$, $1$ or $-1$.}

The first statement is well-known; we omit the easy proof of the second
one. Denote by $A(\zeta)$,

$$
A(\zeta)=[Q(\zeta;\bq)-\alpha]^{-1}
|g_{\bq}(\zeta)\ket \bra g_{\bq}(\zeta)|,
$$
the second term in the representation (\ref{KrOp}) of the
resolvent. Further, denote for $E_0\in\RR$
$$
P_{\alpha}(E_0)=-\Res[R_{\alpha}(\zeta);\zeta=E_0]\,,
$$
$$
P^0(E_0)=-\Res[R^0(\zeta);\zeta=E_0]\,,
$$
$$
T(E_0)=\Res[A(\zeta);\zeta=E_0]\,;
$$
therefore according to (\ref{KrOp})
$$
P_{\alpha}(E_0)=P^0(E_0)+T(E_0)\,.
$$

Start with the proof of the first assertion of Theorem. It is
obvious that $\sigma_1\subset{\rm spec}\,(H_{\alpha})$. Let
$\EE_n\in\sigma_1$, then in a vicinity of $\EE_n$ we have the
following expansion
\begin{equation}
                    \label{Qexp}
Q(\zeta;\bq)-\alpha=
\frac{\partial}{\partial \zeta}Q(\EE_n;\bq)(\zeta-\EE_n)+O(\zeta-\EE_n)^2.
\end{equation}
Therefore
\begin{equation} \label{Qexp1}
T(\EE_n)=\left[\frac{\partial}{\partial\zeta}Q(\EE_n;\bq)\right]^{-1}
|g_{\bq}(\EE_n)\ket\bra g_{\bq}(\EE_n)|\,.
\end{equation}

Since obviously $P^0(\EE_n)=0$, we have
$P_{\alpha}(\EE_n)=T(\EE_n)$ and
the normalized eigenfunction corresponding to $\EE_n$ is
\begin{equation}
                    \label{Qexp2}
\Phi_n=\left[\frac{\partial Q}{\partial \zeta}(\EE_n;\bq)\right]^{-\frac{1}{2}}
g_{\bq}(\EE_n)\,.
\end{equation}

Now consider an eigenvalue $\lambda_n$ of $H^0$. In this case
$P_{\alpha}(\lambda_n)=P^0(\lambda_n)+T(\lambda_n)$. According to
(\ref{QS}), in a neighborhood $W$ of $\lambda_n$ we have the
following representation
$$
g_{\bq}(\zeta)=\Psi_n(\cdot;\bq)(\lambda_n-\zeta)^{-1}+f(\zeta)\,,
$$
where $f$ is analytic function in $W$ with values in $L^2(\RR^3)$
and
$$
\Psi_n(\bx;\bq)=\sum_{k=1}^{k_n}\overline{F_{n,k}(\bq)}F_{n,k}(\bx)\,.
$$

Consider the following three cases: (a) $\lambda_n\in\sigma(\bq)$;
(b) $\lambda_n\notin\sigma(\bq)$ and $Q(\lambda_n;\bq)-\alpha\ne 0$;
(c) $\lambda_n\notin\sigma(\bq)$ and $Q(\lambda_n;\bq)-\alpha=0$.

Let us start with the case (a). Since $\lambda_n$ is a pole of
$Q(\cdot;\bq)$, we have $\Psi_n(\cdot;\bq)\ne0$ and therefore
$T=cP$, where $P$ is the orthoprojector on the one-dimensional
space spanned by $\Psi_n(\cdot;\bq)$. Since $\Psi_n(\cdot;\bq)\in
L_n$, in virtue of statement (B) $c=-1$, and the assertion (2) of
Theorem is proven.

In the case (b) according to Lemma 1,
$F_{n,k}(\bq)=0$ for all $k=1,\ldots,k_n$; hence
$\Psi_n(\cdot,\bq)=0$ and  $T(\lambda_n)=0$. This implies
assertion (3) of Theorem.

Finally, in the case (c) we can use (\ref{Qexp})--(\ref{Qexp2})
with $\zeta=\lambda_n$ instead of $\zeta=\EE_n$, and obtain
$$
T(\lambda_n)=|\Phi_n\ket\bra\Phi_n|\,,
$$
according to (B), this get the statement (4) of Theorem. \halm

\bigskip

For $n\in\NN$ denote by $A_n$ the set of all $\alpha\in\RR$ such
that the solution $\EE_n\equiv\EE_n(\alpha)$ of Equation
(\ref{SE}) does not belong to the spectrum of $H^0$. Lemma 2 shows
that $\RR\setminus A_n$ is finite, moreover, if $V$ bounded from
below, then $A_0=\RR$.

For all $\bq\in\RR^3$ we will denote
$\varepsilon_{-1}(\bq)=\lambda_{-1}=-\infty$. Using Lemmas 1 and 3
we get immediately the following proposition.

\medskip

\noindent{\bf Proposition 1.} {\it For each $n\in\NN$ the function
$\alpha\mapsto \EE_n(\alpha)$ strictly increases on $A_n$.
Moreover},
$$
\lim\limits_{\alpha\to+\infty}\EE_n(\alpha)=\epsilon_n\,,\quad\,\,
\lim\limits_{\alpha\to-\infty}\EE_n(\alpha)=\epsilon_{n-1}\,.\halm
$$

\bigskip

\noindent{\bf Remark.} For $n=0$ we have an interesting phenomenon
of falling the considered particle on the point $\bq$ (the falling
on the attractive center; cf. Ref. \cite{APST} for the case of a
one-dimensional oscillator). Indeed, using estimate (b${}'$) from
Theorem B.7.1 of Ref. \cite{Sim}, we obtain without any difficulty
$|\Phi_0(\bx)|^2\to\delta(\bx-\bq)$ in an appropriate space of
distributions as $\alpha\to-\infty$ (and therefore
$\EE_0\to-\infty$). According to the standard interpretation of
quantum mechanics, this relation means that the probability to
find the particle in a domain not containing the
 point $\bq$ tends to zero as $\EE_0$ tends to $-\infty$.

\section{\label{sec:5}Dependence of the spectrum of $H_\alpha(\bq)$ on $\bq$}

Here we are going to analyze the dependence of the
eigenvalues of $H_\alpha$ on $\bq$. It is clear that $\EE_n(\bq)$
are continuous branches of the multi-valued function defined by
equation (\ref{SE}). This branches can intersect at values
$\lambda_n$ where a monodromy arises. To get a univalent
enumeration of these branches, we modify the parameterization of
the eigenvalues of $H_{\alpha}$ given by Theorem 1 (the
enumeration of the numbers $\EE_n(\bq)$ depends on the enumeration
of poles $\epsilon_n\in{\rm spec}(H^0)$, which in its turn depends
obviously on $\bq$). For $n=-1,0,\ldots$ consider the sets $X_n$
defined as follows: $X_{-1}=\RR^3$, and
$$
X_n=\{\bq\in\RR^3:\exists f\in L_n\,\,\, {\rm s.t.}\,\,\,
f(\bq)\neq 0\}=\{\bq\in\RR^3:\lambda_n\in\sigma(\bq)\}\,,
$$
for $n\ge 0$. For all $n\in\NN$ the set $\RR^3\setminus X_n$ is
nowhere dense in $\RR^3$ (see in Ref. \cite{RS4}, Theorem XIII.63). According
to Lemma 1, for $n\ge0$, the set $X_n$ coincides with the set of
all $\bq\in\RR^3$ such that $\lambda_n$ is a pole of the function
$Q(\cdot;\bq)$. Since we do not suppose the potential $V$ is
smooth, the function $Q(\zeta;\bq)$ on the set
$(\lambda_{n-1},\lambda_n)\times(X_{n-1}\cap X_n)$, $n\ge0$, is
not, generally speaking, smooth. Nevertheless, it is monotone and
real analytic with respect to the first argument $\zeta$ and
continuous with respect to the second argument $\bq$. In this case
the following simple variant of Implicit Function Theorem is
applicable (see in Ref. \cite{Ficht} for the proof):

{\it Let $J$ be an open  nonempty interval of the real line $\RR$,
$X$ be a topological space, and $F:\,J\times X\rightarrow \RR$ be
a separately continuous function such that each partial function
$t\mapsto F(t,x)$, $x\in X$, is strictly monotone. Suppose that
$F(t_0,x_0)=0$ for some $(t_0,x_0)\in J\times X$. Then there are
an open neighborhood $U$ of the point $x_0$ in $X$ and a
continuous function $f:\,U\rightarrow J$ such that $(1)$
$F(f(x),x)=0$ for all $x\in U$; $(2)$ if $U'$ is another
neighborhood of $x_0$, and $g:\,U'\rightarrow J$ is a function with
the property: $F(g(x),x)=0$ for all $x\in U'$, then $U'\subset U$,
and $f_{|U'}=g$. }

According to this version of Implicit Function Theorem, for
any $\bq\in X_{n-1}\cap X_n$ there exists a unique solution
$E_n(\bq)$ to equation (\ref{SE}) that belongs to
$(\lambda_{n-1},\lambda_n)$ and $\bq\mapsto E_n(\bq)$ is a
continuous function in $X_{n-1}\cap X_n$.

\bigskip

\noindent{\bf Proposition 2.}\,
{\it Every function $E_n(\bq)$, $n=0,1,\ldots$,
has a continuous extension to the whole space $\RR^3$.}

\medskip

\noindent{\bf Proof.}\,\, Fix $n=0,1,\ldots$, and let a point
$\bq$, $\bq\in \RR^3\setminus(X_{n-1}\cap X_n)$, be given. Choose
a sequence $(\bq_k)_{k\in\NN}$ from $X_{n-1}\cap X_n$ which tends
to $\bq$. First we note that the sequence $(E_n(\bq_k))_{k\in\NN}$
is bounded in $\RR$. It is trivial for $n>0$. If $n=0$, the
sequence is bounded from above. We prove that it is bounded from
below as well. Otherwise $E_0(\bq_{k_l})\rightarrow-\infty$ for
some subsequence $(\bq_{k_l})$. Since $Q(E;\bq)\rightarrow-\infty$
as $E\rightarrow-\infty$, there exists $A<\lambda_0$ such that
$Q(A;\bq)<\alpha$. Then there exists $N\in\NN$ such that
$Q(A;\bq_{k_l})<\alpha$ and $E_0(\bq_{k_l})<A$ if $l\ge N$.
Therefore, for $k\ge N$ we have
$$
Q(E_0(\bq_{k_l});\bq_{k_l})-\alpha\,<\,Q(A;\bq_{k_l})-\alpha<0\,,
$$
and we get a contradiction with the definition of $E_0(\bq_{k_l})$.

By Bolzano--Weierstrass we can extract a subsequence $(\bq_{k_l})$
from the sequence $(\bq_k)$ such that the subsequence $(E_n(\bq_{k_l}))$
has a limit, which we denote by $E'$. To prove that the sequence
$(E_n(\bq_{k}))$ tends to $E'$ and $E'$ is
independent of the choice of a sequence $(\bq_k)$ tending to $\bq$
we need the following lemma concerning properties of $E'$.

\bigskip

\noindent{\bf Lemma 5.}{\it The limit $E'$ has the properties:
\begin{enumerate}
\item[$(1)$] $E'$ is not a pole of the function $\zeta\mapsto Q(\zeta;\bq)$;
\item[$(2)$] if $\lambda_{n-1}< E'<\lambda_n$, then $E'$ is a unique solution of
Equation  $(\ref{SE})$  in the interval
$(\lambda_{n-1},\lambda_n)$;
\item[$(3)$] if $E'=\lambda_{n-1}$, then
$\displaystyle\lim\limits_{E\to E'}[Q(E;\bq)-\alpha]\geq 0$;
\item[$(4)$] if $E'=\lambda_n$, then
$\displaystyle\lim\limits_{E\to E'}[Q(E;\bq)-\alpha]\leq 0$.
\end{enumerate}
}

\medskip

\noindent{\bf Proof of the lemma.}

(1) First consider the case $n>0$. The function $\tilde
Q_n(\zeta;\bq)=[Q(\zeta;\bq)-\alpha](\zeta-\lambda_{n-1})(\zeta-\lambda_n)$
is continuous on the interval
$(\lambda_{n-2},\lambda_{n+1})\times\RR^3$. Since $\tilde
Q_n(E_n(\bq_{k_l});\bq_{k_l})=0$, passing to the limit
$l\rightarrow\infty$ we get $\tilde Q_n(E';\bq)=0$. Suppose
$\zeta=E'$ is a pole of $Q(\zeta;\bq)$, then $\tilde
Q_n(E';\bq)=\Res[Q(\zeta;\bq);\zeta=E']\neq 0$, and we get a
contradiction. For $n=0$, we consider $\tilde
Q_0(\zeta;\bq)=[Q(\zeta;\bq)-\alpha](\zeta-\lambda_0)$, and get the
same result.

(2) It is sufficient to pass to the limit $l\rightarrow\infty$ in
the identity $Q(E_n(\bq_{k_l});\bq_{k_l})=0$.

(3) In virtue of statement (1) of the lemma, the function
$\zeta\mapsto Q(\zeta;\bq)$ is continuous in a neighborhood of
$E'$, and therefore there exists a limit
$\displaystyle\lim\limits_{\zeta\rightarrow\lambda_{n-1}}[Q(\zeta;\bq)-\alpha]=L$.
Assume that $L<0$, then $Q(E,\bq)-\alpha<0$ for some
$E\in(\lambda_{n-1},\lambda_n)$. Choose some $m$ such that
$E_n(\bq_{k_m})<E$. Since $Q(\zeta;\bq)$ increases on the interval
$(\lambda_{n-1},\lambda_n)$ as the function of $\zeta$, we obtain
a contradiction:
$$
0=Q(E_n(\bq_{k_m});\bq_{k_m})-\alpha<Q(E;\bq_{k_m})-\alpha<0\,.
$$

Statement (4) can be proven similarly to (3).\halm

Let us return to the proof of the proposition. We prove that if a
sequence $(\bp_k)_{k\in\NN}$ from $X_{k-1}\cap X_k$
converges to the point $\bq$, then $E_n(\bp_k)\rightarrow E'$.

Suppose $E_n(\bp_k)$ does not converge to $E'$, then there exists
a subsequence $(\bp_{k_l})$ such that $E_n(\bp_{k_l})\rightarrow
E^*$, $E^*\ne E'$. Assume $E^*<E'$. Taking into account item (2)
of Lemma 5 we get $E^*=\lambda_{n-1}$ or $E'=\lambda_n$. In both
the cases we have
$$
\lim_{\zeta\rightarrow E^*}[Q(\zeta;\bq)-\alpha]\geq 0\;\;\textrm{and}\;\;
\lim_{\zeta\rightarrow E'}[Q(\zeta;\bq)-\alpha]\leq 0.
$$
Take some real numbers $E_1$ and $E_2$ such that $E^*<E_1<E_2<E'$.
Then by the strict monotonicity of $\zeta\mapsto Q(\zeta;\bq)$ we have
$$
0\leq Q(E_1;\bq)-\alpha<Q(E_2,\bq)-\alpha\leq 0\,.
$$
This is a contradiction. \halm

\bigskip

The following theorem is the main result of this section.

\bigskip

\noindent{\bf Theorem 2.} {\it
For each fixed $\alpha\in\RR$ there is a sequence $(E_n(\bq))_{n\in\NN}$ of
continuous functions of $\bq\in\RR^3$ with the following properties:
\begin{enumerate}
\item[$(1)$] $\lambda_{n-1}\leq E_n(\bq)\leq \lambda_n$ for all $n\in\NN$.
\item[$(2)$] For each $\bq\in\RR^3$ the set consisting of all
$E_n(\bq)$ and all the numbers $\lambda_n$ with multiplicities
$k_n>1$ form the complete collection of the eigenvalues of the
operator $H_\alpha(\bq)$.
\item[$(3)$] If $\lambda_{n-1}<E_n(\bq)<\lambda_n$,
then $E_n(\bq)$ is a unique solution of the equation
$($\ref{SE}$)$ on the interval $(\lambda_{n-1},\lambda_n)$.
\item[$(4)$] If $\zeta=\lambda_n$ is a pole of the function $\zeta\mapsto Q(\zeta;\bq)$, then
$E_{n-1}(\bq)<\lambda_n<E_n(\bq)$.
\item[$(5)$] If $\zeta=\lambda_n$ is not a pole of the function $\zeta\mapsto Q(\zeta;\bq)$,
then we have the following assertions:
\begin{enumerate}
\item[$(a)$] if $Q(\lambda_n;\bq)-\alpha<0$, then $E_n(\bq)=\lambda_n<E_{n+1}(\bq)$;
\item[$(b)$] if $Q(\lambda_n;\bq)-\alpha>0$, then $E_n(\bq)<\lambda_n=E_{n+1}(\bq)$;
\item[$(c)$] if $Q(\lambda_n;\bq)-\alpha=0$, then $E_n(\bq)=\lambda_n=E_{n+1}(\bq)$.
\end{enumerate}
\end{enumerate}
}

\medskip

\noindent{\bf Proof.} Consider the functions $E_n(\bq)$ given
by  Proposition 2. Then (1) is obvious by definition of $E_n(\bq)$.
Assertion (2) follows from Theorem 1. Assertions (3) and (4) were
proven in Lemma 5. It remains to prove (5).

Let $\lambda_n$ be not a pole of $\zeta\mapsto Q(\zeta;\bq)$.
Suppose $Q(\lambda_n;\bq)-\alpha<0$. For any positive integer $m$
we choose a number $E'_m$ such that
$\lambda_n-1/m<E'_m<\lambda_n$; then
$Q(E'_m;\bq)-\alpha<0$. Further, we choose points $\bq_m\in\RR^3$
such that $\lambda_{n-1}$ and $\lambda_n$ are not poles of the
function $\zeta\mapsto Q(\zeta;\bq_m)$ (that is $\bq_m\in
X_{m-1}\cap X_m$), and such that $\displaystyle
|\bq-\bq_m|<1/m$ and $Q(E'_m;\bq_m)-\alpha<0$. Then
$\zeta=E_n(\bq_m)$ is a solution of the equation
$Q(\zeta;\bq_m)-\alpha=0$ lying in the interval
$(\lambda_{n-1},\lambda_n)$. Since $Q(\zeta;\bq_m)$ is a strictly
monotone function of $\zeta$ on this interval, the inequalities
$E'_m<E_n(\bq_m)<\lambda_n$ take place for all $m$. Thus
$E_n(\bq_m)\rightarrow\lambda_n$ and $\bq_m\rightarrow\bq$ as
$m\rightarrow\infty$; therefore $\lambda_n=E_n(\bq)$ by the
definition of the function $E_n(\bq)$. According to Lemma 5,
$Q(\lambda_n;\bq)-\alpha\geq 0$, if $\lambda_n=E_{n+1}(\bq)$;
therefore  $\lambda_n<E_{n+1}(\bq)$. Hence, item (5a) is proved.
The proofs of items (5b) and (5c) are similar. \halm

\bigskip

Theorem 2 gives a useful description of the spectrum of
$H_\alpha$. Namely, denote by $M$ the set $\{m\in\NN:\,k_m>1\}$
and together with the functions $E_n(\bq)$ introduce a sequence of
constant functions $\Lambda_m^{(k)}(\bq)=\lambda_m$, where $m\in
M$, $k=1,\ldots,k_m-1$. Then $E_n(\bq)\le\Lambda_n^{(k)}(\bq)\le
E_{n+1}(\bq)$ for all $n\in M$, $k=1,\ldots,k_n-1$, and for any
fixed $\bq\in\RR^3$ the union of the sequences
$(E_n(\bq))_{n\in\NN}$ and $(\Lambda_m^{(k)}(\bq))_{m\in M,
k=1,\ldots,k_m-1}$ forms the complete set of the eigenvalues of
$H_{\alpha}(\bq)$ multiplicity counting. If
$\displaystyle\bq\in\bigcap\limits_{n=0}^{\infty}X_n$, then every
$E_n(\bq)$ is distinct from the numbers $\Lambda_m^{(k)}(\bq)$.
Since $\displaystyle\RR\setminus\bigcap\limits_{n=0}^{\infty}X_n$
is the set of the first Baire category, for a generic $\bq$ the
point perturbation levels $E_n(\bq)$ are distinct from the levels
of the unperturbed operator $H^0$.

\section{\label{sec:6}Point perturbations of the harmonic oscillator}

Here we apply the results of the previous sections to
the Hamiltonian (\ref{OP}) with the potential
\begin{equation}
                           \label{AHO}
V(\br)=\frac{\mu\Omega_x^2}{2}x^2+
\frac{\mu\Omega_y^2}{2}y^2+\frac{\mu\Omega_z^2}{2}z^2\,,
\end{equation}
where $\Omega_j$ ($j=x,y,z$) are the frequencies of the
oscillator. The function $V$ can be considered as a confinement
potential of a quantum well in $\RR^3$ with the characteristic
sizes
$$
L_j=\sqrt{\frac{\hbar}{2\mu\Omega_j}}\,,\quad\quad j=x,y,z,
$$
(numbers $\sqrt{2}L_j$ are called also {\it length parameters} of
the oscillator\textsuperscript{\cite{ED}}). Therefore the operator with potential
(\ref{AHO}) can be used as the Hamiltonian of a (generally
speaking, asymmetric) quantum dot.\textsuperscript{\cite{BGL}} It is convenient to
pass to dimensionless coordinates $\bx=\br/L$, where
$L=\sqrt[3]{L_xL_yL_z}$. In the coordinates $\bx=(x_1,x_2,x_3)$
the operator $\hat H^0$ takes the form $\hat H^0=\hbar\Omega H^0$,
where
$$
H^0=-\Delta+\frac{1}{4}\left(\omega_1^2 x_1^2+
\omega_2^2 x_2^2+\omega_3^2 x_3^2\right)\,,
$$
$$
\Omega=\sqrt[3]{\Omega_x\Omega_y\Omega_z}\,,\quad
\omega_1=\frac{\Omega_x}{\Omega}\,,\quad
\omega_2=\frac{\Omega_y}{\Omega}\,,\quad
\omega_3=\frac{\Omega_z}{\Omega}\,
$$
(hence, $\omega_1\omega_2\omega_3=1$).

Further we discuss the properties of $H^0$. The spectrum of this
operator consists of the eigenvalues
$$
\lambda_{n_1 n_2
n_3}=\omega_1(n_1+1/2)+\omega_2(n_2+1/2)+\omega_3(n_3+1/2)\,,
$$
where $n_1,n_2,n_3\in\NN$.
The corresponding normalized eigenfunctions are
$$
\Phi_{n_1n_2n_3}(\bx)=
\phi_{n_1}(x_1)\phi_{n_2}(x_2)\phi_{n_3}(x_3)\,,
$$
where
$$
\phi_{n_j}(x_j)=\left(\frac{\omega_j}{2\pi}\right)^{1/4}(2^{n_j}n_j!)^{-1/2}
\exp\left(-\frac{1}{4}\omega_j x_j^2\right)
H_{n}\left(\sqrt{\frac{\omega_j}{2}}x_j\right)
$$
is the oscillator function
($H_n(x)$ is the Hermite polynomial of degree $n$).

If the frequencies $\omega_1,\omega_2,\omega_3$ are independent
over the ring $\ZZ$ (this is the generic case), then the spectrum
of $H^0$ is simple; therefore the multiplicity of the eigenvalues
of $H_{\alpha}(\bq)$ does not exceed 2 and the part $\sigma_2$ of
the spectrum ${\rm spec}\,(H_{\alpha}(\bq))$ is always empty. On
the other hand, since $H_n(0)=0$ if and only if $n$ is odd,
$\lambda_{n_1,n_2,n_3}\in {\rm spec}(H_\alpha(0))$ if and only if
one of the numbers $n_j$ ($j=1,2,3$) is odd; hence, ${\rm
spec}(H^0)\setminus\sigma(0)$ is always infinite. In addition, for
all $n>0$ the set $\RR^3\setminus X_n$ is infinite.

In general case, there are no explicit expressions for the Green
functions of the harmonic oscillator in terms of commonly used
elementary or special functions. Nevertheless, in a number of
cases, the representation of the Green function $G^0(\bx,\by;E)$
as the Laplace transform of the heat kernel $K(\bx,\by;t)$ for
$H^0$ is very useful to investigate some properties of the Krein
$\mathcal{Q}$-function. The heat kernel for $H^0$ has the form
(see, e.g., in Ref. \cite{FH}):
$$
K^0(\bx,\by;t)=\prod\limits_{j=1}^3\left(\frac{1}{4\pi\sh\omega_jt}\right)^{1/2}
\exp\left(-\frac{\omega_j}{4\sh\omega_jt}((x_j^2+y_j^2)\ch\omega_jt-2x_jy_j)\right)\,.
$$
Using the heat kernel $K^f$ for the free Hamiltonian
$H^f=-\Delta$,
$$
K^f(\bx,\by;t)=(4\pi t)^{-3/2}
\exp\left(-\frac{(\bx-\by)^2}{4t}\right)\,,
$$
and the $\mathcal{Q}$-function for $H_f$,
$$
Q^f(\zeta)=-\frac{\sqrt{-\zeta}}{4\pi}\,,
$$
we get immediately from the formula
$$
G(\bx,\by;E)=\int\limits_0^{\infty}e^{tE}K(\bx,\by;t)\,dt\,,
$$
that for ${\rm Re}\,\zeta
<(\omega_1+\omega_2+\omega_3)/2$ the following
representation of the $\mathcal{Q}$-function for $H^0$ takes
place:
$$
Q(\zeta;\bq)=-\frac{\sqrt{-\zeta}}{4\pi}
$$
\begin{equation}
                     \label{QAHO}
+\frac{1}{(4\pi)^{3/2}}
\int\limits_0^{\infty}\left(\prod\limits_{j=1}^3\left(\frac{1}{\sh\omega_jt}\right)^{1/2}
\exp\left(-\frac{1}{2}{q_j^2\omega_j\tgh\displaystyle
\frac{\omega_jt}{2}}\right)-\frac{1}{t^{3/2}}\right)\,e^{\zeta
t}\,dt\,.
\end{equation}

It is clear from (\ref{QAHO}) that $(\dd Q/\dd
q_j)(E;\bq)<0$ for $q_j>0$, if $E<\lambda_0=(\omega_1+\omega_2+\omega_3)/2$. Since
$\dd Q/\dd E>0$ for $E\in\RR\setminus {\rm
spec}\,(H^0)$, (\ref{SE}) implies that $\dd E_0/\dd q_j>0$. 
In particular, the depth of the lowest
impurity level $\lambda_0- E_0(\bq)$ decreases if $|\bq|$
increases in such a way that the inner product $\ba\cdot\bq$ remains
positive for each vector $\ba$ with positive coordinates. In the
spherically symmetric case $\omega_1= \omega_2=\omega_3$, we have
$\dd Q/\dd q<0$, where $q=|\bq|>0$, and the
depth decreases with increasing of $q$. This phenomenon was
discovered numerically for a spherically symmetric quantum dot 
in Ref. \cite{KZ}
and called {\it positional disorder}. We see that the
positional disorder is common to each parabolic quantum dot, not
only to the spherically symmetric one. The similar result is valid
in the two-dimensional case, i.e. for the case of impurities in a
quantum well (see numerical results in Ref. \cite{KZ}). Our arguments
are valid in the two-dimensional case also, thus we have a strict
proof for the positional disorder in a two-dimensional quantum
well.

The more detailed analysis is possible in the case of the {\it
isotropic oscillator}: $\Omega_x=\Omega_y=\Omega_z$ $(=\Omega)$,
i.e. in the case of a spherically symmetric quantum dot. In this
case $\omega_1=\omega_2=\omega_3=1$ and the spectrum of $H^0$
consists of the eigenvalues
$$
\lambda_n=n+\frac{3}{2}\,,\quad\,n\in\NN\,,
$$
where $\lambda_n$ has the multiplicity $k_n=(n+1)(n+2)/2$. 
In this case there are natural units of
length (namely, $L$) and of energy ($\hbar\Omega$). Therefore the
following very important scaling properties  takes place. Denote
by $\hat Q(\zeta;\bq)$ the Krein $\Q$-function for the operator
$\hat H^0$ keeping the notation $Q(\zeta;\bq)$ for the
$\Q$-function of $H^0$. Then
$$
\hat Q(\zeta;\bq)=\frac{1}{\hbar\Omega
L^3}Q\left(\frac{\zeta}{\hbar\Omega};\frac{\bq}{L}\right)=
4\pi\frac{\mu}{2\pi\hbar^2
L}Q\left(\frac{\zeta}{\hbar\Omega};\frac{\bq}{L}\right)\,.
$$
Denote $\mu/(2\pi\hbar^2 L)$ by $\alpha^0$;
obviously, $\alpha^0$ is strength of the point potential
corresponding to the scattering length $L$. Then Equation
(\ref{SE}) takes the form
\begin{equation}
                         \label{scalSE}
4\pi Q\left(\frac{\zeta}{\hbar\Omega};\frac{\bq}{L}\right)=
\frac{\alpha}{\alpha^0}\,,
\end{equation}
or, equivalently,
$$
4\pi Q\left(\frac{\zeta}{\hbar\Omega};\frac{\bq}{L}\right)=
\frac{L}{\ell_s}\,.
$$
Equation (\ref{scalSE}) shows that a change of the frequency
$\Omega$ does not change the numerical values of energy levels in
the spectrum of $\hat H^0$ if  $L$ is used as the unit of length,
$\hbar\Omega$ as the unit of energy and $\alpha^0$ as the unit
of point potential strength.

In the  case of isotropic oscillator, the set $\sigma(\bq)$ has a
simple description:

\bigskip

\noindent{\bf Proposition 3.} {\it Let
$\Omega_x=\Omega_y=\Omega_z$. Then
$\sigma(\bq)=\{\lambda_{2n}:\,n\in\NN\}$, if $\bq=0$, and
$\sigma(\bq)= {\rm spec}(H^0)$ otherwise}.

\medskip

\noindent{\bf Proof.}\,\,\, Each $\lambda_n$ is equal to
$\lambda_{n_1n_2n_3}$, where $n_1+n_2+n_3=n$. If $n$ is odd, then
at least one of $n_j$ is odd, and $\Psi_{n_1n_2n_3}(0)=0$.
Therefore $\lambda_n\notin\sigma(0)$. On the other hand, if $n$ is
even, then $\Psi_{n00}(0)\ne0$, and therefore $\lambda_n\in
\sigma(0)$.

Let now $\bq\ne0$. First we remark that for all $n\in\NN$ the
following assertion is valid:

\noindent{\bf Lemma 6}.
 {\it If $H_n(x_0)=0$, then $H_{n+1}(x_0)\ne 0$}.

\medskip

\noindent{\bf Proof of the lemma.}\,For all $n\in \NN$ the
following relation takes place:\textsuperscript{\cite{BE}}
$$
H'_{n+1}(x)=2(n+1)H_n(x)\,.
$$
If $H_n(x_0)=H_{n+1}(x_0)=0$, then $H'_n(x_0)=0$. Since $y=H_n(x)$
is a solution to the differential equation $y''-2xy'+2ny=0$, we
have $H_n(x)=0$ for all $x$; but this is impossible.  \halm

\medskip

Let us return to the proof of the proposition. Suppose that
$\bq\ne 0$; without loss of generality we can assume $q_2\ne 0$.
Since $H_1(x)=0$ only for $x=0$, and $H_0(x)\ne0$ for all $x$, we
have $\lambda_0,\lambda_1\in\sigma(\bq)$. Let $n>1$. Suppose that
$\Phi_{n-1,1,0}(\bq)=0$, then according to Lemma~6,
$\Phi_{n,0,0}(\bq)\ne0$. \halm

Using Proposition 3 we can give the complete description of the
spectrum $H_\alpha(\bq)$ in the case of an isotropic $H^0$.
Moreover, in this case the explicit form of the Green function
$G^0(\bx,\bx';\zeta)$ is known, and therefore we can give the explicit
form of the Krein $\Q$-function and eigenfunction of
$H_\alpha(\bq)$. In particular, the equation for the point
perturbation levels $E_n(\bq)$ can be obtained in an explicit
form. The mentioned Green function has the form:\textsuperscript{\cite{GS}}
$$G^0(\bx,\by;\zeta)=-\frac{1}{2(2\pi)^\frac{3}{2}}
\Gamma\left(\frac{1}{2}-\zeta\right)
\Bigg[
\frac{U(-\zeta;\xi)U'(-\zeta;-\eta)+U'(-\zeta;\xi)U(-\zeta;-\eta)}{|\bx-\by|}$$
\begin{equation}\label{G0}
+\frac{U(-\zeta;\xi)U'(-\zeta;-\eta)-U'(-\zeta;\xi)U(-\zeta;-\eta)}{|\bx+\by|}\Bigg],
\end{equation}
where $\xi=(|\bx+\by|+|\bx-\by|)/2$,
$\eta=(|\bx+\by|-|\bx-\by|)/2$, $U(\nu;z)$ is the
parabolic cylinder function\textsuperscript{\cite{AS}} (in the Whittaker notation
$U(\nu;z)=D_{-\nu-1/2}(z)$), and $U'$ denotes the derivative of
$U$ with respect to the second argument
$$
U'(\zeta;y)=\frac{\dd}{\dd y}U(\zeta;y)\,.
$$

Using (\ref{G0}), we get the following expression for the
$Q$-function:
$$
Q(\zeta;\bq)=
-\frac{1}{8(2\pi)^{3/2}}\Gamma\left(\frac{1}{2}-\zeta\right)
$$
$$
\times\big[
\left(q^2-4\zeta\right)U(-\zeta,q)U(-\zeta,-q)+4U'(-\zeta,q)U'(-\zeta,-q)
$$
\begin{equation}\label{QF}
-\frac{2}{q}\big(U'(-\zeta,q)U(-\zeta,-q)-U(-\zeta,q)U'(-\zeta,-q)\big)\big],
\end{equation}
where $q=|\bq|$. Due to the symmetry of the problem, the
$\Q$-function depends on $q$ only, so we shall write often
$Q(\zeta;q)$ instead of $Q(\zeta;\bq)$. Introducing the notation
$\Uu(\zeta;y)=U(\zeta;y)U(\zeta;-y)$, we can rewrite (\ref{QF}) in
the sometimes more useful form
\begin{equation}\label{Q2}
Q(\zeta;\bq)=-\frac{1}{4(2\pi)^{\frac{3}{2}}}
\Gamma\left(\frac{1}{2}-\zeta\right)\bigg[
\left(q^2-4\zeta\right)\Uu(-\zeta;q)- \frac{1}{q}\Uu\,'(-\zeta;q)-
\Uu\,''(-\zeta;q)\bigg]\,,
\end{equation}
where the prime denotes the derivative with respect to the second
argument as before. Passing to limit we get at $q=0$
\begin{equation}
              \label{Q0}
Q(\zeta;0)=-\frac{1}{\sqrt{8}\pi}
\frac{\displaystyle \Gamma\left(\frac{3}{4}-\frac{\zeta}{2}\right)}
{\displaystyle \Gamma\left(\frac{1}{4}-\frac{\zeta}{2}\right)}\,.
\end{equation}
It is interesting to compare (\ref{Q0}) with the Krein
$\Q$-function $Q^{(1)}(\zeta;0)$ for the one-dimensional harmonic
oscillator:\textsuperscript{\cite{GC97}}
$$
Q^{(1)}(\zeta;0)=2^{-3/2} \frac{\displaystyle
\Gamma\left(\frac{1}{4}-\frac{\zeta}{2}\right)} {\displaystyle
\Gamma\left(\frac{3}{4}-\frac{\zeta}{2}\right)}\,.
$$
Curiously, in the case of the free Hamiltonian $H^0=-\Delta$, the
$\mathcal{Q}$-functions $Q_d$ for $d=1$ and for $d=3$ are also related
as follows:
\begin{equation}
                                 \label{Cur}
Q_1^{-1}(\zeta)=-8\pi Q_3(\zeta)\,.
\end{equation}
Namely, for the free Hamiltonian
$Q_1(\zeta)=(2\sqrt{-\zeta})^{-1}$,
$Q_3(\zeta)=-(4\pi)^{-1}\sqrt{-\zeta}$. For $q\ne 0$ relation
(\ref{Cur}) for $\mathcal{Q}$-functions of the harmonic
oscillators is violated.

It is useful to consider the behavior of the function $\zeta
\mapsto Q(\zeta;q)$ near the singular points, i.e., near the poles
and in a neighborhood of $-\infty$. Using properties of the
parabolic cylinder functions,\textsuperscript{\cite{BE}} we have
$$
Q(\zeta;0)=-\frac{(2n+1)!!}{(2\pi)^\frac{3}{2}(2n)!!}\left(\frac{1}{\zeta-\lambda_{2n}}-\ln
2+1-\frac{1}{2}\sum_{k=1}^n\frac{1}{k(1+2k)}
+O(\zeta-\lambda_{2n})\right),
$$
as $\zeta\to \lambda_{2n}$.
If $\bq\neq 0$, the coefficients for corresponding asymptotics are
cumbrous enough, and we give the leading term only:
$$
Q(\zeta;q)=-\frac{\exp\left(-q^2/2\right)}{(2\pi)^{3/2}2^{n+2}n!}
\left( 2(n+1) H^2_n\left(q/\sqrt{2}\right)\right.$$
\begin{equation}\label{QA}
\left.+\sqrt{2}\left(q^{-1}-q\right)H_n\left(q/\sqrt{2}\right)
H_{n+1}\left(q/\sqrt{2}\right)+H^2_{n+1}\left(q/\sqrt{2}\right)
\right)(\zeta-\lambda_n)^{-1}+O(1)\,,
\end{equation}
as $\zeta\to\lambda_{n}$.

For ${\rm Re}\,\zeta\to-\infty$, we have
\begin{equation}
                 \label{As1}
Q(\zeta;q)=-\frac{\sqrt{-\zeta}}{4\pi}
\left(1-\frac{q^2}{8}\zeta^{-1}+\frac{8-q^4}{128}\zeta^{-2}+
O(\zeta^{-3})\right)\,.
\end{equation}
It is important to note that the leading term in (\ref{As1})
coincides with the Krein ${\cal Q}$-function for the free
Hamiltonian $-\Delta$.

Now consider the properties of the function $q\mapsto Q(\zeta;q)$.
Since $U(\nu;z)$ is an entire function of $z$, the function
$q\mapsto Q(\zeta;q)$ at $\zeta\notin{\rm spec}(H^0)$ can be
extended to a real analytic even function on $\RR$ (see
(\ref{Q2})). In particular,
$$
\frac{\dd}{\dd q}Q(\zeta;0)=0\,.
$$
As to the second derivative, we can obtain after some algebra
\begin{equation}
                           \label{derQ2}
\frac{\partial^2}{\partial q^2}Q(\zeta;0)=
\frac{1}{8\sqrt{6}\pi}\left[
\left(4\zeta^2+1\right)
\frac{\Gamma(\frac{1}{4}-\frac{\zeta}{2})}
{\Gamma(\frac{3}{4}-\frac{\zeta}{2})}
-8\zeta\frac{\Gamma(\frac{3}{4}-
\frac{\zeta}{2})}{\Gamma(\frac{1}{4}-\frac{\zeta}{2})}
\right].
\end{equation}
For the fixed $\zeta\in\RR\setminus{\rm spec}(H^0)$, the
asymptotics of $Q$ at $q\to\infty$ is given by
\begin{equation}
                    \label{asQ}
Q(\zeta;q)=-\frac{1}{8\pi}
\left[q-\frac{2\zeta}{q}-\frac{1+2\zeta^2}{q^3}+
O\left(\frac{1}{q^5}\right)\right]\,.
\end{equation}
This follows from the asymptotics for $\Uu(\zeta;q)$ at
$q\to\infty$:\textsuperscript{\cite{Mil}}
$$
\Uu(\zeta;q)=\frac{\sqrt{2\pi}}{\Gamma(\frac{1}{2}+\zeta)}
\bigg[\frac{1}{X}+O\bigg(\frac{1}{X^5}\bigg)\bigg]\,,
$$
where $X=\sqrt{q^2+4\zeta}$.

\begin{figure}
\resizebox{\hsize}{!}{\includegraphics*{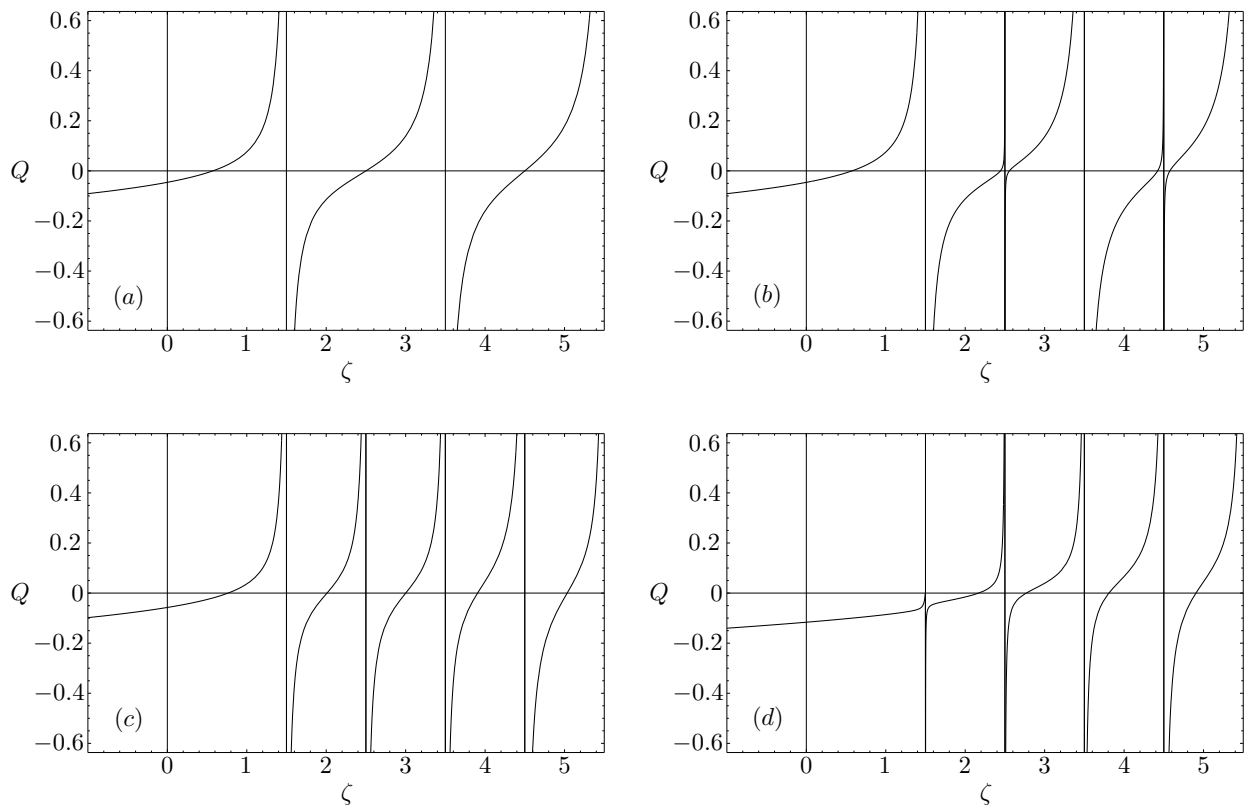}}
\caption{
$Q$ as a function of $\zeta$ for
$(a)\;q=0,\;(b)\;q=1/10,\;(c)\;q=1,\;(d)\;q=3$.
}\label{fig:1}\end{figure}

\begin{figure}
\resizebox{\hsize}{!}{\includegraphics*{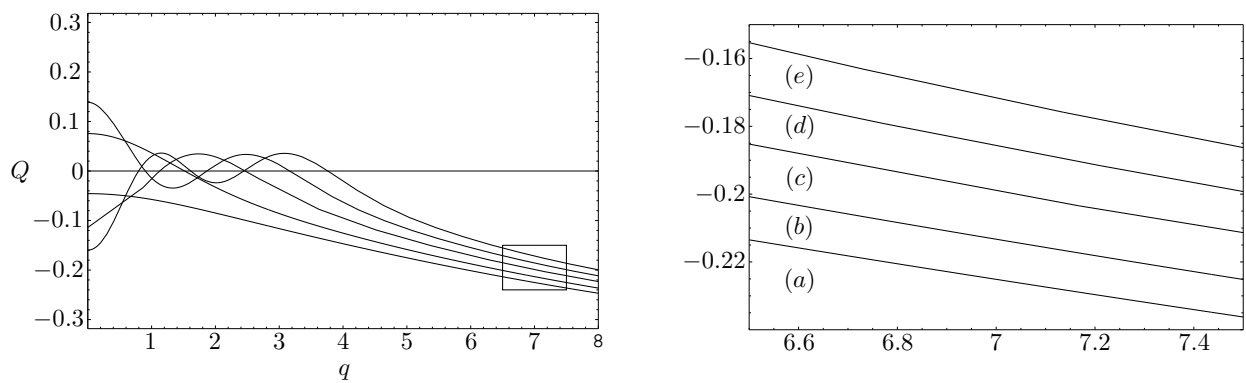}}
\caption{
$Q$ as a function of $q$ for
$(a)\;\zeta=0,\;(b)\;\zeta=1,\;(c)\;\zeta=2,\;(d)\;\zeta=3,\;(e)\;\zeta=4$.
}\label{fig:2}\end{figure}

Further the following formula will be also useful
\begin{equation}
\label{derQ0}
\frac{\partial Q}{\partial \zeta}(\zeta;0)=
\frac{1}{4\sqrt{2}\pi}
\frac{\Gamma(\frac{3}{4}-\frac{\zeta}{2})}
{\Gamma(\frac{1}{4}-\frac{\zeta}{2})}
G\left(\frac{1}{2}-\zeta\right)\,.
\end{equation}
Here and below we use the standard notations\textsuperscript{\cite{BE}}
$$
G(z)=\psi\left(\frac{z}{2}+\frac{1}{2}\right)-\psi\left(\frac{z}{2}\right)\,;\quad\quad
\psi(z)=\frac{\Gamma'(z)}{\Gamma(z)}\,.
$$
The plot of the graphs for
the function $Q(\zeta;q)$ is shown on Fig. \ref{fig:1} and Fig. \ref{fig:2}.

In the case of an isotropic oscillator, the functions $E_n(\bq)$
depend only on $q$ and we will denote them by $E_n(q)$. Further
properties of these functions (and, in particular, of the spectrum
of $H_\alpha(\bq)$) for the isotropic case are given in Theorem 3
below, which is one of the main results of the article.

\medskip

\noindent{\bf Theorem 3.}{\it The following assertions take
place.}

\begin{itemize}
\item[(1)] {\it The functions $E_n(q)$, $n\in\NN$ are real-analytic.
If $\alpha=0$ and $n>0$, then in a vicinity of zero, these functions are continuous
branches of a two-valued analytic function.}

\item[(2a)] {\it $E_0(0)<\lambda_0$ for each $\alpha$, $\alpha\in\RR$}.

 \item[(2b)] {\it If $\alpha>0$, then  $E_{2n+1}(0)=\lambda_{2n+1}$ and
 $\lambda_{2n+1}<E_{2n+2}(0)<\lambda_{2n+2}$
 $\forall n\in\NN$}.

 \item[(2c)] {\it If $\alpha<0$, then
$\lambda_{2n}<E_{2n+1}(0)<\lambda_{2n+2}$ and $E_{2n+2}(0)=\lambda_{2n+2}$
$\forall n\in\NN$}.

 \item[(2d)] {\it If $\alpha=0$, then
$E_{2n+1}(0)=E_{2n+2}(0)=\lambda_{2n+1}$ $\forall n\in\NN$}.

\item[(3a)] {\it If $\alpha\ne0$, then for any  $n>0$}
\begin{equation}
                    \label{der1}
\frac{\partial E_n}{\partial q}(0)=0\,.
\end{equation}
\noindent{\it If $n=0$, then} (\ref{der1}) {\it is valid for any
$\alpha$}.

\item[(3b)]{\it If $\alpha>0$ $($respectively, $\alpha<0)$, then}
\begin{equation}
                    \label{der2}
\frac{\partial^2 E_{n}}{\partial q^2}(0)=
\frac{1}{8\sqrt{6}G(\frac{1}{2}-E_{n}(0))}\left(
\frac{4E_{n}^2(0)+1}{8\pi^2\alpha^2}
-8E_{n}(0)\right)
\end{equation}
\noindent{\it for any even $($respectively, odd $)$ $n$.
If $n=0$, then} (\ref{der2}) {\it is valid for any
$\alpha$}.

\item[(3c)] {\it If $\alpha=0$, then $(\dd E_{2n+1}/\dd q)(0)<0$,
$(\dd E_{2n+2}/\dd q)(0)>0$, and
$(|(\dd E_{2n+1}/\dd q)(0)|=|(\dd E_{2n+2}/\dd q)(0)|$
 $\forall n\in\NN$}.

\item[(4)] {\it If $q\ne0$, then $\lambda_{n-1}<E_n(q)<\lambda_n$
$\forall n\in\NN$}.

\item[(5)]
$\lim\limits_{q\to\infty}E_n(q)=\lambda_n$
$\forall n\in\NN$.
\end{itemize}

\medskip

\noindent{\bf Proof.}\,\,\,Item (4) follows immediately from
Proposition 3 and the definition of the functions $E_n$. Formula
(\ref{Q0}) shows that $Q(\zeta;0)=0$ if and only if
$\zeta=\lambda_{2n+1}$ for some $n\in\NN$; therefore items
(2a)--(2d) follow from Theorem 2. Using the standard version of
the implicit function theorem and the Proposition 3 again, we see
that $E_n(q)$ are real analytic at $q>0$. Moreover, item (3) of
Theorem 2 implies that (i) $E_n(q)$ are real-analytic at $q=0$ for
even $n$ if $\alpha>0$, (ii) $E_n(q)$ are real-analytic at $q=0$
for odd $n$ if $\alpha<0$, and (iii) $E_0(q)$ is real-analytic at
$q=0$ for any $\alpha$. In all these cases, the derivatives of
$E_n$ can be found from the equations
\begin{equation}
                        \label{QQ1}
\frac{\partial Q}{\partial \zeta}\frac{\partial E_n}{\partial q}+
\frac{\partial Q}{\partial q}=0\,,
\end{equation}
$$
\frac{\partial Q}{\partial\zeta}\frac{\partial^2E_n}{\partial q^2}+
\frac{\partial^2 Q}{\partial \zeta^2}\left(\frac{\partial E_n}{\partial q}\right)^2
+2\frac{\partial^2 Q}{\partial \zeta\partial q}\frac{\partial E_n}{\partial q}+
\frac{\partial^2 Q}{\partial q^2}=0.
$$

Since $(\partial Q/\partial q)(E;0)=0$ if
$E\notin {\rm spec}(H^0)$, equation (\ref{der1}) follows from
(\ref{QQ1}) in the considered cases. In virtue of (\ref{der1}),
the second derivative of $E_n$ is given by
\begin{equation}
                          \label{QQ3}
\frac{\partial^2 E_n}{\partial q^2}(0)= -\frac{\partial^2
Q}{\partial q^2} \left(\frac{\partial Q}{\partial
\zeta}\right)^{-1}(E_n(0);0).
\end{equation}
Substituting (\ref{derQ2}) and  (\ref{derQ0}) in (\ref{QQ3})
and using (\ref{SE}) we get (\ref{der2}).

Now consider the singular case when $E_n(0)$, $n\ge1$, coincides
with a point of the form $\lambda_{2m+1}$. In a neighborhood of
the point $(E_n(0),0)$, introduce the function
$$
\tilde Q_{\alpha}(\zeta;q)=\frac{Q(\zeta;q)-\alpha}{\displaystyle
\Gamma\left(\frac{1}{2}-\zeta\right)}\,,
$$
which is smooth with respect to $(\zeta,q)$ and analytic with respect to
the first argument $\zeta$. In a vicinity of $(E_n(0),0)$ we have
\begin{equation}
                          \label{QQ4}
\tilde Q_{\alpha}(E_n(q);q)=0\,.
\end{equation}
Further,
\begin{equation}
                          \label{QQ5}
\frac{\dd \tilde Q_{\alpha}}{\dd
\zeta}=\frac{1}{\Gamma(1/2-\zeta)}\frac{\dd Q}{\dd
\zeta}+(Q-\alpha)\frac{\Gamma'(1/2-\zeta)}{\Gamma^2(1/2-\zeta)}\,.
\end{equation}
Since $(\dd Q/\dd\zeta)(\zeta;0)$ is a finite
number at $\zeta=\lambda_{2m+1}$ and $\Gamma(1/2-\zeta)$ has a
pole at $\lambda_{2m+1}$, the first term in (\ref{QQ5}) vanishes
at the point $(\lambda_{2m+1},0)$. The value of the function
$\Gamma'(1/2-\zeta)/\Gamma^2(1/2-\zeta)$ at $\zeta=\lambda_{2m+1}$
is a nonzero finite number. Finally, $Q(\lambda_{2m+1},0)=0$; thus
$\dd \tilde Q_{\alpha}/\dd q$ vanishes at the point
$\lambda_{2m+1}$ if and only if $\alpha=0$. Therefore, if
$\alpha\ne 0$, then each function $E_n(q)$ has an analytic
continuation in a neighborhood of the point $q=0$. Since $q\mapsto
\tilde Q_{\alpha}(\lambda_{2m+1};q)$ is an even function, we get
easily (\ref{der1}).

Let now $\alpha=0$. Then
$$
\frac{\dd^2 \tilde Q_0}{\dd \zeta^2}=
 \frac{1}{\Gamma(1/2-\zeta)}\frac{\dd^2 Q}{\dd
\zeta^2}
$$
\begin{equation}
                                       \label{QQ6}
+2\frac{\Gamma'(1/2-\zeta)}{\Gamma^2(1/2-\zeta)}\frac{\dd Q}{\dd
\zeta}-Q\frac{\Gamma''(1/2-\zeta)\Gamma(1/2-\zeta)-2\Gamma'^2(1/2-\zeta)}
{\Gamma^3(1/2-\zeta)}\,.
\end{equation}
It is easy to see that the first and last terms in (\ref{QQ6})
vanishes at the point $(\lambda_{2m+1},0)$, whereas the second one
does not. Therefore, $\dd^2 \tilde Q_0/\dd\zeta^2\ne0$ at the
point $(\lambda_{2m+1},0)$, and $E_n(q)$ being solutions of
(\ref{QQ4}), are continuous branches a two-valued analytic
function in a vicinity of $(\lambda_{2m+1},0)$. Obviously,
at the point $(\lambda_{2m+1},0)$ the following relation is valid
$$
\frac{\dd^2 \tilde Q_0}{\dd \zeta^2}\left(\frac{\dd E_n}{\dd
q}\right)^2+2\frac{\dd^2 \tilde Q_0}{\dd \zeta\dd q}\frac{\dd
E_n}{\dd q} +\frac{\dd \tilde Q_0}{\dd \zeta}\frac{\dd^2 E_n}{\dd
q^2}+\frac{\dd^2 \tilde Q_0}{\dd q^2}=0\,.
$$
Since $\dd \tilde Q_0/\dd \zeta=0$ at the
considered point, we get the quadratic equation for
$\dd E_n/\dd q$:
$$
\frac{\dd^2 \tilde Q_0}{\dd \zeta^2}\left(\frac{\dd E_n}{\dd
q}\right)^2+\frac{\dd^2 \tilde Q_0}{\dd q^2}=0\,.
$$
As a result, we complete the proof of items (1) and (3c). It
remains to prove (5). Fix $n\in\NN$ and let $\epsilon$,
$0<\epsilon<1$, is given. According to (\ref{asQ}) we can choose
$q_0>0$ such that $Q(\lambda_n-\epsilon;q)-\alpha<0$ if $q\ge
q_0$. Since $Q(E_n(q);q)-\alpha=0$ and the function $E\mapsto
Q(E;q)$ increases in the interval $\lambda_{n-1}<E<\lambda_n$, we
have $E_n(q)>\lambda_n-\epsilon$ as $q\ge q_0$. Moreover,
$E_n(q)<\lambda_n$, and the proof is completed. \halm

\begin{table}\center
\begin{tabular}{c|c|c}
\hline
\hline 
 &$q=0$&$q\neq0$\\\hline
 $\alpha>0$& $\sigma_1=\{E_{2n}(0):\,n\in\NN\}$ & $\sigma_1=\{E_n(q):\,n\in\NN\}$ \\
 & $\sigma_2=\{\lambda_{2n+2}:\,n\in\NN\}$ & $\sigma_2=\{\lambda_{n+1}:\,n\in\NN\}$ \\
 & $\sigma_3=\{\lambda_{2n+1}:\,n\in\NN\}$ & $\sigma_3=\emptyset$ \\
 & $\sigma_4=\emptyset$ & $\sigma_4=\emptyset$ \\ 
 \hline
 $\alpha=0$ & $\sigma_1=\{1/2\}$ & $\sigma_1=\{E_n(q):\,n\in\NN\}$ \\
 & $\sigma_2=\{\lambda_{2n+2}:\,n\in\NN\}$ & $\sigma_2=\{\lambda_{n+1}:\,n\in\NN\}$ \\
 & $\sigma_3=\emptyset$ & $\sigma_3=\emptyset$ \\
 & $\sigma_4=\{\lambda_{2n+1}:\,n\in\NN\}$ & $\sigma_4=\emptyset$ \\ 
 \hline  
 $\alpha<0$ & $\sigma_1=\{E_{2n+1}(0):\,n\in\NN\}\cup\{E_0(0)\}$ & $\sigma_1=\{E_n(q):\,n\in\NN\}$ \\
 & $\sigma_2=\{\lambda_{2n+2}:\,n\in\NN\}$ & $\sigma_2=\{\lambda_{n+1}:\,n\in\NN\}$ \\
 & $\sigma_3=\{\lambda_{2n+1}:\,n\in\NN\}$ & $\sigma_3=\emptyset$ \\
 & $\sigma_4=\emptyset$ & $\sigma_4=\emptyset$ \\ 
\hline\hline\end{tabular}
\caption{The structure of ${\rm spec}(H_\alpha(q))$.}
\label{tab:1}
\end{table}

The structure of ${\rm spec}(H_\alpha(q))$ given by Theorem 3
is presented in Table \ref{tab:1}.
The peculiarities of this table at $q=0$ can be understood from
the point of view the symmetry group of the problem. It is well
known that for a generic spherically symmetric potential $V(\br)$,
the eigenvalues $\lambda$ of the operator $H^0=-\Delta+V$ are
parameterized by three quantum numbers:
$\lambda=\lambda_{n_r,l,m}$, where $n_r$ ($n_r=0,1,\ldots$) is the
so called principal (or total) quantum number; $l$
($l=0,1,\ldots$) is the orbital quantum number, and $m$
($m=-l,-l+1,\ldots,l-1,l$) is the magnetic quantum number. Each
eigenvalue $\lambda_{n_r,l,m}$ is degenerate with multiplicity
$2l+1$, namely, $\lambda_{n_r,l,m}=\lambda_{n_r,l,m'}$ if
$m,m'\in\{-l,-l+1,\ldots,l-1,l\}$. This degeneracy is related to
the invariance of $H^0$ with respect to the rotation group
$\SM\OO(3)$: eigensubspaces of $H^0$ carry an irreducible
representation of this group. In general, $\lambda_{n_r,l,m}\ne
\lambda_{n_r',l',m'}$ if $n_r\ne n_r'$ or $l\ne l'$. The
eigenvalues of an isotropic harmonic oscillator have an additional
(so-called accidental) degeneracy: each eigensubspace $L_n$ is
decomposed on the subspaces $L_n^{(l)}$ with angular momentum
$l=n,n-2,\ldots,0$ (if $n$ is even) or $l=n,n-2,\ldots,1$ (if $n$
is odd). This accidental degeneracy is related to the invariance
of the Hamiltonian $H^0$ of an isotropic harmonic oscillator with
respect to the group $\UU(3)$. Indeed,
$$
H^0=\sum\limits_{j=1}^3a_j^+a_j+\frac{3}{2}\,,
$$
where $a_j^+$ and $a_j$ are standard creation and annihilation
operators.\textsuperscript{\cite{ED}} Therefore, $H^0$ is invariant with respect to
the transformation
$$
a_j\to a'_j=\sum\limits_{j=1}^3u_{kj}a_j\,,\quad\quad
a^+_j\to a'^+_j=\sum\limits_{j=1}^3u^*_{kj}a^+_j\,,
$$
where $(u_{jk})$ is a unitary matrix. If $q=0$, then $H_\alpha(0)$
is a spherically symmetric perturbation of $H^0$ that violates the
$\UU(3)$-symmetry. To prove this, we note that operators
$a^+_ja_k$ are generators of the Lie group $\mathfrak{u}(3)$.
Therefore, if $H_\alpha(0)$ is invariant with respect to the
considered representation of $\UU(3)$, we must have
$[H_\alpha(0),H^0]$=0. On the other hand it is easy to show that
for $\zeta\in\CC\setminus\RR$ the operator
$[R_\alpha(\zeta),R^0(\zeta)]$ has a nonzero integral kernel.

Since  point perturbations can not change states with nonzero
angular momentum $l$ (see, e.g., \cite{AGHH}), the part $\sigma_2$ (at $q=0$)
may contain only even eigenvalues $\lambda_{2n}$ and we see this
in Table \ref{tab:1}. Since all states from $L_n$ have the same parity
$(-1)^n$, the isotropic oscillator has no stationary states with a
nonzero dipole momentum.\textsuperscript{\cite{BZP}} On the other hand every
eigensubspace of $H_0(0)$ with eigenvalue from $\sigma_4$ have an
eigenfunction with $l=0$ (this is the eigenfunction from item 4 of
Theorem 1). Therefore point perturbations of an isotropic
harmonic oscillator can lead to an appearance of eigenstates with
nonzero dipole momentum.

An alternative tool to understand the energy degeneracy of the
three-dimensional iso\-tro\-pic oscillator gives the supersymmetry
theory.\textsuperscript{\cite{KNT}, \cite{LRB1}, \cite{LRB2}} 
We will not dwell
here on this approach, nevertheless note that the analysis
performed in the cited papers requires a modification in the
$s$-channel only.

The functions $E_n$ depend not only on the position parameter $q$,
but also on the strength $\alpha$; we will denote these
dependencies as $E_n=E_n(q,\alpha)$. If $E(q,\alpha_0)$ coincides
with one of the numbers $\mathcal{E}_m$, then in a vicinity of
$\alpha_0$, the function $\alpha\mapsto E_n(q,\alpha)$ is a
continuous branch of the inverse function to $E\mapsto Q(E;q)$. It
is already known from Proposition 1 that the following limits take
place:
$$
\lim_{\alpha\to+\infty} E_n(q;\alpha)=\lambda_n,\;\;\;\;
\lim_{\alpha\to-\infty}E_n(q;\alpha)=\lambda_{n-1},
$$
where $\lambda_{-1}=-\infty$. Now we make more precise this
behavior. From (\ref{As1}) we get the asymptotics of the function
$E_0(q;\alpha)$ for the fixed $q\geq0$ as $\alpha\to-\infty$,
\begin{equation}
                            \label{as1}
E_0(q;\alpha)=-16\pi^2\alpha^2+\frac{q^2}{4}+\frac{1}{128\pi^2\alpha^2}+O\left(\frac{1}{\alpha^4}\right)\,,
\end{equation}
or in terms of the point perturbation of the initial operator
(\ref{HOS})
\begin{equation}
                \label{as2}
E_0(q;\alpha)=-\frac{\hbar^2}{2\mu
l_s^2}+\frac{\mu\Omega^2q^2}{2}+\frac{\mu\Omega^2l_s^2}{4}+
O(l_s^4)\,,
\end{equation}
where the scattering length $l_s$ tends to $0$. Expression
(\ref{as2}) means that up to the infinitely small term $O(l_s^2)$
the ground state of $\hat H_\alpha(\bq)$ equals to the ground state
of the point perturbation of the free Hamiltonian
$-\hbar^2\Delta/2\mu$ with the same scattering length $l_s$
shifted by the potential $V(\br)=\mu\Omega^2\br^2/2$ at the point
$\br=\bq$. Equation (\ref{as2}) shows that at least for the
isotropic harmonic oscillator its potential can be recovered from
the dependence of the ground state of the point perturbation on
the position of the potential support. It is reasonable to suppose
that this is true for more general forms of the potential $V$; we
consider this conjecture elsewhere.

Now consider the behavior of $E_n(q;\alpha)$ in a vicinity of the
poles of $Q(\zeta,q)$. We start with the general case $q\ne 0$.
Using (\ref{QA}) we get as $\alpha\to\pm\infty$
$$
E_n(q;\alpha)=\lambda_n^{\pm}
-\frac{\exp(-q^2/2)}{(2\pi)^{3/2}2^{n+2}n!} \left(2(n+1)
H^2_n\left(q/\sqrt{2}\right)\right.
$$
\begin{equation}\label{EA}
\left.+\sqrt{2}\left(q^{-1}-q\right)H_n\left(q/\sqrt{2}\right)
H_{n+1}\left(q/\sqrt{2}\right)+H^2_{n+1}\left(q/\sqrt{2}\right)
\right)\alpha^{-1}+O(\alpha^{-2}),
\end{equation}
where $\lambda_n^+=\lambda_n$ and $n\ge0$ as $\alpha\to+\infty$,
and $\lambda_n^-=\lambda_{n-1}$ and $n\ge 1$ as
$\alpha\to-\infty$.

\begin{figure}
\resizebox{\hsize}{!}{\includegraphics*{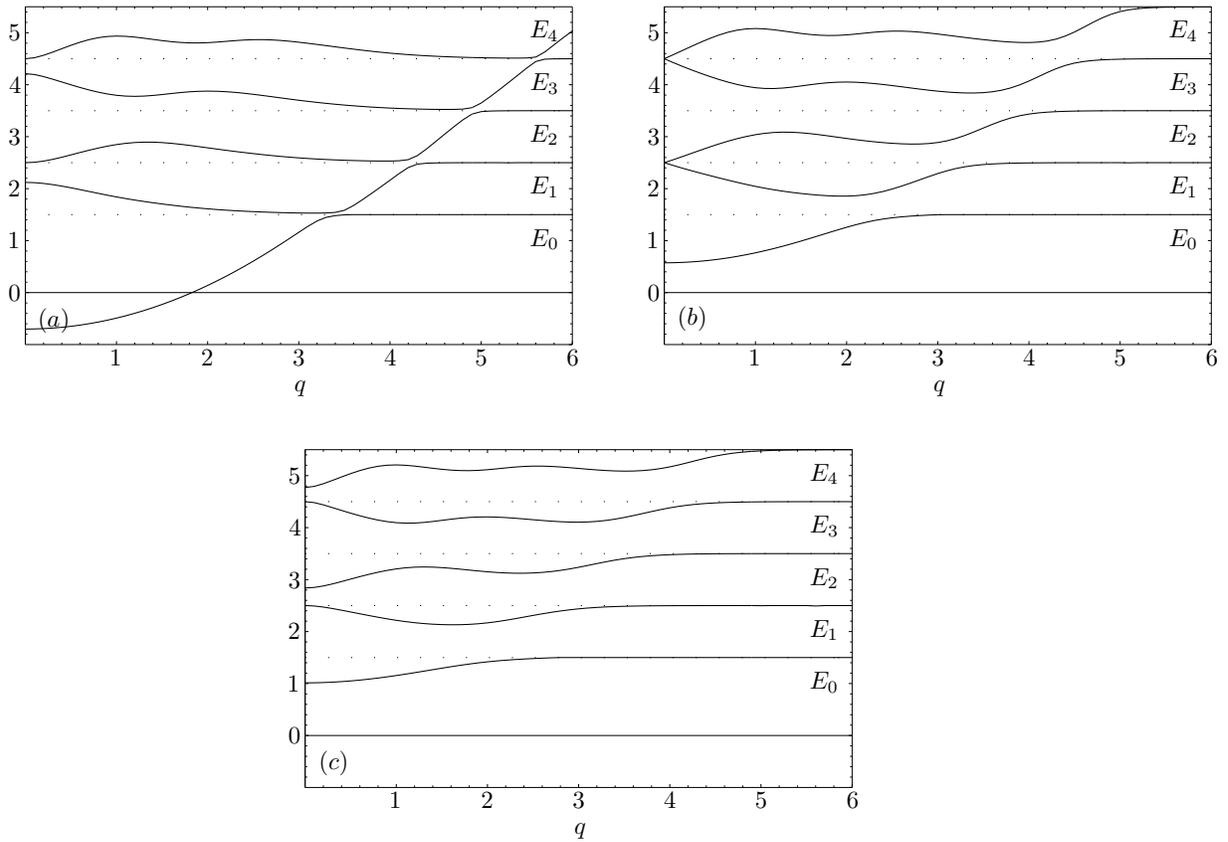}}
\caption{
$E_n$ as a function of $q$ for $(a)\;\alpha=-\alpha^0,$
$(b)\;\alpha=0,$ $(c)\;\alpha=\alpha^0$.
}\label{fig:3}\end{figure}

\begin{figure}
\resizebox{\hsize}{!}{\includegraphics*{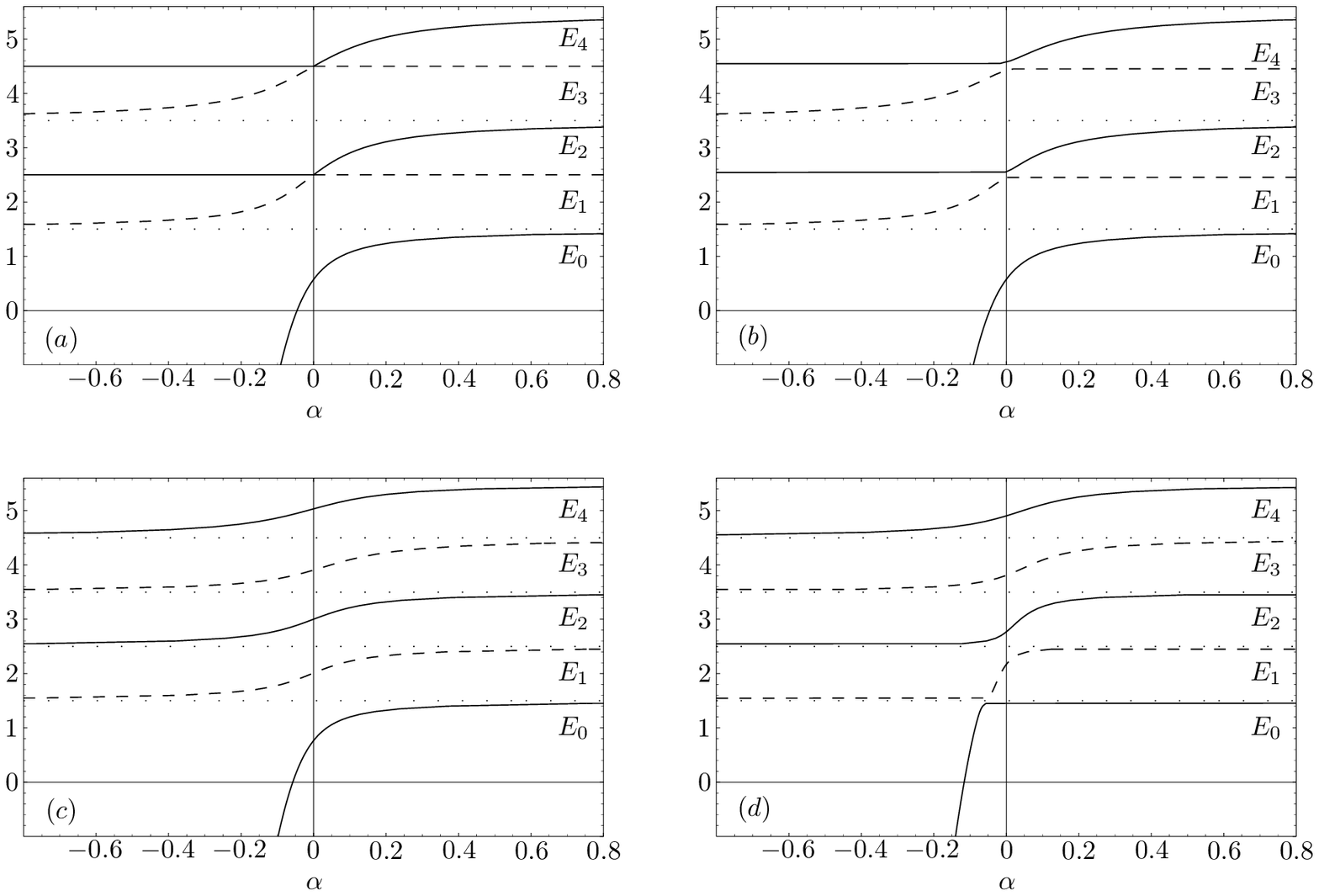}}
\caption{
$E_n$ as a function of $\alpha$ for
$\displaystyle(a)\;q=0,(b)\;q=1/10,(c)\;q=1,(d)\;q=3$.
}\label{fig:4}\end{figure}

In the case $q=0$, we are in position to give a compact form for
more precise asymptotics of $E_n(q;\alpha)$. Denote
$$
\Lambda_n(\alpha)=
\frac{(2n+1)!!}{(2\pi)^\frac{3}{2}(2n)!!}\alpha^{-1}
-\left(\frac{(2n+1)!!}{(2\pi)^\frac{3}{4}(2n)!!}\right)^2\left(\ln
2-1+\frac{1}{2}\sum_{k=1}^n\frac{1}{k(1+2k)}\right)\alpha^{-2}.
$$
For eigenvalues with even indices we have
\begin{equation}\label{EA0}
E_{2n}(0;\alpha)=\left\{
\begin{array}{ll}
\lambda_{2n-1} & \textrm{for $\alpha\le 0$ and $n\ge1$}, \\
\noalign{\bigskip} \displaystyle -16\pi^2\alpha^2+
\frac{1}{128\pi^2}\alpha^{-2}+O(\alpha^{-4})&
\textrm{for $\alpha\to-\infty$ and $n=0$}, \\
\noalign{\bigskip} \displaystyle
\lambda_{2n}-\Lambda_n(\alpha)+O(\alpha^{-3})
 & \textrm{for $\alpha\to+\infty$ and $n\ge0$}.
\end{array}\right.
\end{equation}
For the odd indices
\begin{equation}\label{EA1}
E_{2n+1}(0;\alpha)=\left\{
\begin{array}{ll}
 \lambda_{2n+1} & \textrm{for $\alpha\ge 0$}, \\
\noalign{\bigskip} \displaystyle
\lambda_{2n}-\Lambda_n(\alpha)+O(\alpha^{-3})
& \textrm{for $\alpha\to-\infty$}.
\end{array}\right.
\end{equation}
Formulas (\ref{EA}) -- (\ref{EA1}) explain peculiarities in the
plots of functions $E_n$ on Figures \ref{fig:3} and \ref{fig:4}. Note that in
Eqs.~(\ref{as1}) -- (\ref{EA1}) the remainder terms depend on $n$.

The isotropic harmonic oscillator has an equidistant spectrum.
After the perturbation by a zero-range potential, the distances
between energy levels are changed and become dependent on the
energy index $n$. This is important in the connection with the
problem of the controlled modulation of the binding energy of the
impurity center in quantum dots, that can be used to design
nonlinear opto-electronic active elements.\textsuperscript{\cite{KGAS}}
The asymptotic formulas (\ref{as2}) -- (\ref{EA1})
give very accurate expressions for the excited energies in the
most interesting case of a deep zero-range well
($\alpha\to-\infty$) as well as for the case of a shallow well
($\alpha\to+\infty$), which confirm numerical results from Ref.
\cite{KGAS}. Note also that Proposition 1 and Theorem 3 imply a
remarkable distinction between the excited energy for the ground
state and that for the other ones: the energy
$E_1(q;\alpha)-E_0(q;\alpha)$ can take an arbitrary value
depending on $q$ and $\alpha$; on the other hand, energies
$\lambda_n-E_n(q;\alpha)$ and $E_{n+1}(q;\alpha)-\lambda_n (n\ge
1)$ are bounded by $1$. Since at fixed $\alpha$, $\alpha\ll-1$,
the function $q\mapsto E_1(q;\alpha)-E_0(q;\alpha)$ is injective
for moderate values of $q$, the position of an impurity in the
quantum dot may be determined from the spectroscopy data.

\begin{figure}
\resizebox{\hsize}{!}{\includegraphics*{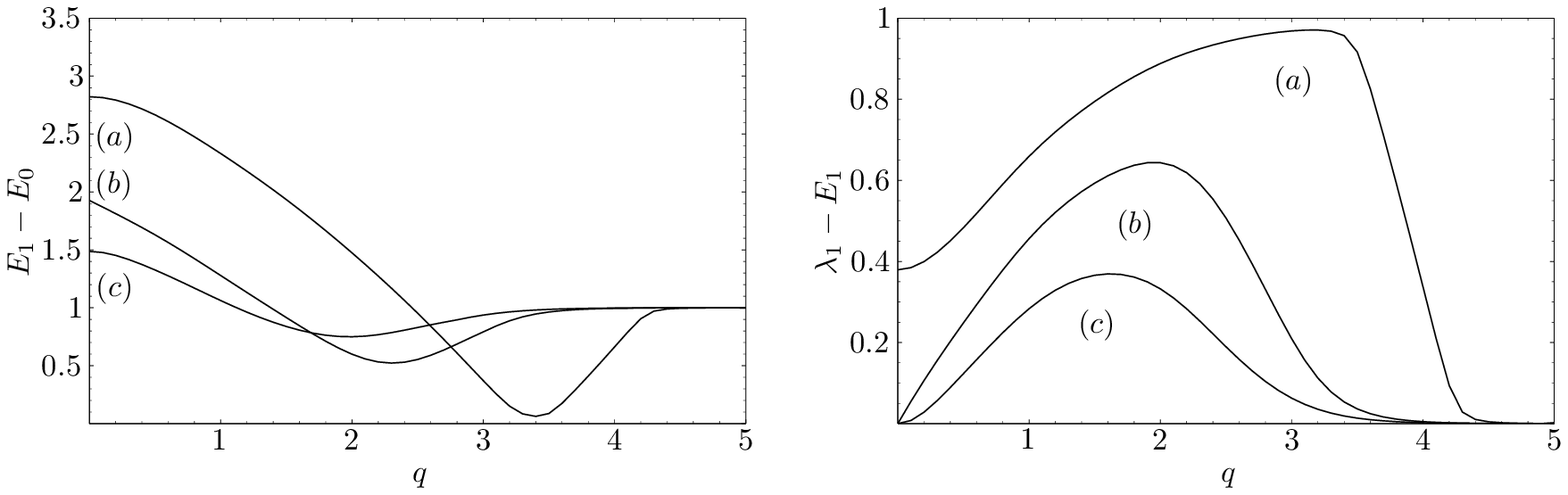}}
\caption{
The exciting energy as a function of $q$ for
$(a)\;\alpha=-\alpha^0,\;(b)\;\alpha=0,\;
(c)\;\alpha=\alpha^0$.
}\label{fig:5}\end{figure}

\begin{figure}
\resizebox{\hsize}{!}{\includegraphics*{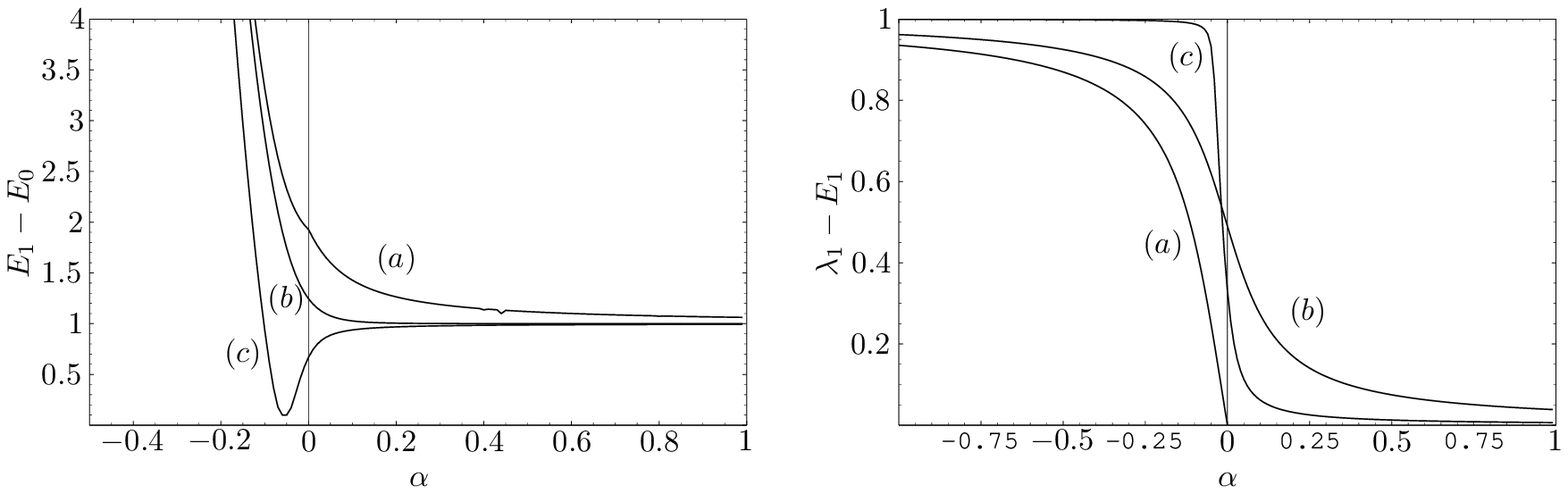}}
\caption{
The exciting energy as a function of $\alpha$ for
$(a)\;q=0,\;(b)\;q=1,\;(c)\;q=3$.
}\label{fig:6}\end{figure}

We show  the plot of the energies
$E_1(q;\alpha)-E_0(q;\alpha)$ and $\lambda_1-E_1(q;\alpha)$ as
functions of $q$ and $\alpha$ on Fig.~\ref{fig:5} and Fig.~\ref{fig:6}, respectively.

In conclusion we give the following remark.
Let $q:\,[0,\infty)\to \RR$ be a smooth function obeying the
conditions

\begin{itemize}

\item[(H1)] {\it $q\ge0$ and the function $r\mapsto q(r)+r^2/4$ is
nondecreasing},

\item[(H2)] $q'(r)\le 0$,

\end{itemize}

\noindent and let $\kappa_0$ and $\kappa_1$ are the first two
eigenvalues of the operator $H^0+q=-\Delta+r^2/4+q(r)$. It is
proven in Ref. \cite{Kur} that $\kappa_0/\kappa_1<\lambda_0/\lambda_1$,
if $q\ne0$. Using Theorems 3 and A it is easy to construct smooth
functions $q$ with properties (H1) and

\begin{itemize}

\item[(H2a)] $q'(r)\ge 0$,

\end{itemize}

\noindent such that $\kappa_0/\kappa_1>\lambda_0/\lambda_1$.

\section*{ACKNOWLEDGMENTS}
This work was partially supported by SFB-288, INTAS (Grant 00-257),
DFG (Grant 436 RUS 113/572/0-2), and RFBR (Grant 02-01-00804).

\setlength{\labelsep}{0pt}


\begin{thebibliography}{99}

\bibitem{BGL} D.~Bimberg, M.~Grundmann, and N.~N.~Ledentsov,
        {\it Quantum dot heterostructures}
        (J.~Wiley \& Sons, Chichester etc., 1999).

\bibitem{Har} P.~Harrison,
        {\it Quantum wells, wires and dots}
        (J.~Wiley \& Sons, Chichester etc., 2000).

\bibitem{LUSKAB} N.~N.~Ledentsov, V.~M.~Ustinov, V.~A.~Shchukin., P.~S.~Kop'ev,
       Zh.~I.~Alferov, and D.~Bimberg,
       ''Quantum dot heterostructures: fabrication, properties, lasers
       (Review) (in Russian),'' 
       Fiz. Tekh. Poluprovodn. {\bf 32}, 385-410 (1998);
       Engl. transl. Sov. Phys. Semicond. {\bf 32}, 343-365 (1998).

\bibitem{QD} {\it Quantum dots,} edited by 
        E.~Borovitskaya and M.~S.~Shur 
        (World Sci., New Jersey etc., 2002).

\bibitem{FBP} J.~M.~Ferreyra, P.~Bosshard, and C.~R.~Proetto,
         ''Strong-confinement approach for impurities in parabolic quantum dots,''
         Phys. Rev. B. {\bf 55}, 13682-13688 (1997).

\bibitem{KGAS} V.~D.~Krevchik, A.~B.~Grunin, A.~K.~Aringazin, and M.~B.~Semenov,
       ''Magneto-optical properties of the quantum dot - impurity
       center systems synthesized in a transparent dielectric matrix,'' 
       {\tt cond-mat/0309011}.

\bibitem{LGPW} E.~Lind, B.~Gustafson, I.~Pitzonk, and L.-E.~Wernersson,
       ''Tuneling spectroscopy of a quantum dot through a single impurity,''
       Phys. Rev. B. {\bf 68}, 033312, 1-4 (2003).

\bibitem{AGHH}  S.~Albeverio, F.~Gesztesy, R.~H{\o}egh-Krohn, and H.~Holden,
        {\it Solvable models in quantum mechanics}
        (Springer--Verlag, Berlin etc., 1988).

\bibitem{Pav} B.~S.~Pavlov,
        ''The theory of extensions and explicily-solvable models,''
        Russ.~Math.~Surv. {\bf 42}, 127-186 (1987).

\bibitem{Dav} J.~H.~Davies,
         {\it The physics of low-dimensional semiconductors}
         (Cambridhe Univ. Press, Cambridge, 1998).

\bibitem{BS} C.~Bose and C.~K.~Sarkar,
        ''Effect of a parabolic potential on the impurity binding energy
        in spherical quantum dots,''
        Physica~B {\bf 253}, 238-241 (1998).

\bibitem{KPS} E.~M.~Kazaryan,  L.~S.~Petrosyan, and  H.~A.~Sarkisyan,
        ''Impurity states in a narrow band gap semiconductor quantum dot with
        parabolic confinement potential,''
        Physica E {\bf 16}, 174-178 (2003).

\bibitem{KZE} V.~D.~Krevchik, R.~V.~Zaitsev, and V.~V.~Evstifeev,
        ''On the theory of photoionization of deep-level impurity centers in a
        parabolic quantum well,'' 
        Semiconductors {\bf 34}, 1193-1198 (2000).

\bibitem{KZ} V.~D.~Krevchik and R.~V.~Zaitsev,
        ''Impurity absorption of light in structures with quantum dots,''
        Phys. Solid States {\bf 46}, 522-526 (2001).

\bibitem{KGZ} V.~D.~Krevchik, A.~B.~Grunin, and R.~V.~Zaitsev,
       ''Anisotropy of the magneto-optical absorption of quantum dot -
       impurity center complex,'' 
       Semiconductors {\bf 36}, 1146-1153 (2002).

\bibitem{MF} T.~Martin and S.~Feng,
       ''Suppression of scattering in electron transport in mesoscopic quantum
       Hall systems,''
       Phys. Rev. Lett. {\bf 64}, 1971-1974 (1990).

\bibitem{BVK} V.~L.~Bakhrakh, S.~I.~Vetchinkin, and S.~V.~Khristenko,
        ''Green's function of a multidimensional isotropic harmonic oscillator,''
        Theor. Math. Phys. {\bf 12}, 776-778 (1972).

\bibitem{GS} C.~Grosche and F.~Steiner,
        {\it Handbook of Feynman path integrals}
        (Springer-Verlag, Berlin etc., 1998).

\bibitem{KM} D.~B.~Khrebtukov and J.~H.~Macek,
        ''Harmonic oscillator Green functions,''
        J. Phys. A: Math. Gen. {\bf 31}, 2853-2868 (1998).

\bibitem{FI96} S.~Fassari and G.~Inglese,
        ''Spectroscopy of a three-dimensional isotropic harmonic oscillator with a
        $\delta$-type perturbation,''
        Helv.~Phys.~Acta {\bf 69}, 130-140 (1996).

\bibitem{APST} M.~P.~Avakian, G.~S.~Pogosian, A.~N.~Sissakian, and V.~M.~Ter-Antonyan,
        ''Spectroscopy of a singular linear oscillator,''
        Phys.~Lett.~A {\bf 124}, 233-236 (1987).

\bibitem{PZ} V.~V.~Papoyan and V.~A.~Zagrebnov,
        ''On condensation of a one-dimensional nonideal boson gas,''
        Phys.~Lett. {\bf 113A}, 8-10  (1985).

\bibitem{FI94} S.~Fassari and G.~Inglese,
        ''On the spectrum of the harmonic oscillator with a $\delta$-type
        perturbation,''
        Helv.~Phys.~Acta {\bf 67}, 650-659 (1994).

\bibitem{FI97} S.~Fassari and G.~Inglese,
        ''On the spectrum of the harmonic oscillator with a $\delta$-type perturbation. II,''
        Helv.~Phys.~Acta {\bf 70}, 858-865 (1997).%

\bibitem{PWZ} Q.-Z.~Peng, X.~Wang, and J.-Y.~Zeng,
        ''Analytic solution to the Schr\"odinger equation with a harmonic oscillator
        potental plus $\delta$-potential,''
        Sc.~in~China~(Ser.~A.) {\bf 34}, 1215-1221 (1991).

\bibitem{GC97} V.~A.~Geyler and I.~V.~Chudaev,
        ''The spectrum of a quasi-two-dimensional system in a parallel
        magnetic field  (in Russian),''
        Zhurn.~Vychislit.~Matematiki~i~Matem.~Fiziki {\bf 37}, 214-222 (1997);
        Engl.~transl. Comput.~Math.~and~Math.~Phys. {\bf 37}, 210-218 (1997).

\bibitem{GC98} V.~A.~Geyler and I.~V.~Chudaev,
        ''Schr\"odinger operators with moving point perturbations and related solvable models
        of quantum mechanical systems,''
        Z.~f\"ur~Analysis~und~Anwend. (J.~for~Analisis~and~Appl.)
        {\bf 17}, 37-55 (1998).

\bibitem{BG} J.~Br\"uning and V.~Geyler,
        ''Scattering on compact manifolds with infinitely thin horns,''
        J.~Math.~Phys. {\bf 44}, 371-405 (2003).

\bibitem{CFKS} H.~L.~Cycon, R.~G.~Fr\"ose, W.~Kirsch, and B.~Simon,
          {\it Schr\"odinger operators with applications to qauntum mechanics and global
          geometry} 
          (Springer-Verlag, Berlin etc., 1987).

\bibitem{GMC1} V.~A.~Geyler, V.~A.~Margulis, and I.~I.~Chuchaev,
        ''Zero-range potentials and Carleman operators  (in Russian),''
        Sibirsk.~Matem.~Zhurn. {\bf 36}, 828-841 (1995);
        Engl.~transl. Siberian~Math.~J. {\bf 36}, 714-726 (1995).

\bibitem{Sim} B.~Simon,
        ''Schr\"odinger semigroups,''
        Bull.~Amer.~Math.~Soc. {\bf 7}, 447-526 (1982).

\bibitem{RS2} M.~Reed and B.~Simon,
        {\it Methods of modern mathematical physics. II: Fourier analysis, self-adjointness}
        (Academic~Press, New York etc., 1975).

\bibitem{Mir} C.~Miranda,
       {\it Equazioni alle derivate parziali di tipo ellittico}
       (Springer-Verlag, Berlin, 1955).

\bibitem {Ber} Yu.~Berezanski,
        {\it Expansion in eigenfunctions of self-adjoin operators}
        (AMS, Providence, RI, 1968).

\bibitem{BZP} A.~I.~Baz', Y.~B.~Zeldovich, and A.~M.~Perelomov,
        {\it Scattering, reactions and decay in nonrelativistic quantum mechanics}.
        (Israel Progr. Ssi. Transl., Jerusalem, 1969).

\bibitem{DO} Yu.~N.~Demkov and V.~N.~Ostrovskii,
         {\it Zero-range potentials and their applications in atomic
         physics} 
         (Plenum Press, New York, 1988).

\bibitem{KL} M.~G.~Krein and H.~K.~Langer,
        ''Defect subspace and generalized resolvents of an Hermitian
        operators in the space $\Pi_{\kappa}$ (in Russian),''
        Funk. Anal. i Prilozh. {\bf 5}, 59-71 (1971);
        Engl.~transl. Funct. Anal. and Appl {\bf 5}, 217-228 (1971).

\bibitem{RS4} M.~Reed and B.~Simon,
        {\it Methods of modern mathematical physics. IV: Analysis of operators}
        (Academic~Press, New York etc., 1978).

\bibitem{DM} V.~A.~Derkach and M.~M.~Malomud,
        ''Generalized resolvents and the boundary value
        problems for Hermitian operators with gaps,''
        J.~Funct.~Anal. {\bf 95}, 1-95 (1991).

\bibitem{Ficht} G.~M.~Fichtenholz,
        {\it Differential- und Integralrechnung. I}
        (Dt.~Verl. der~Wiss., Berlin, 1972). 

\bibitem{ED} J.~P.~Elliott and P.~G.~Dawber,
        {\it Symmetry in  physics, V.}~II
        (McMillan Press: London, 1979).

\bibitem{FH} R.~P.~Feynman and A.~R.~Hibbs,
        {\it Quantum mechanics and path integrals}
        (McGraw-Hill: New York, 1965).

\bibitem{BE} H.~Bateman and A.~Erdelyi,
        {\it Higher transcendental functions} 
        (Mc Graw-Hill book Comp., New York etc., 1981).

\bibitem{AS} M.~Abramowitz and I.~A.~Stegun,
         {\it Handbook of mathematical functions with formulas, graphs, and mathematical
         tables} 
         (John Wiley {\&} Sons, New York, 1984).

\bibitem{Mil} J.~C.~P.~Miller,
       {\it Tables of Weber parabolic cylinder functions.
       Giving solutions of the differential equation
       $d^2y/dx^2+(x^2/4-a)y=0$.
       Computed by Scientific Computing Service Limited.}
       (Her Majesty's Stationary office, London, 1955).

\bibitem{KNT}  V.~A.~Kostelecky, M.~M.~Nieto, and D.~R.~Traux,
        ''Supersymmetry and the relationship between the Coulomb and oscillator
        ''problems in arbitrary dimensions,''
        Phys. Rev. D. {\bf 32}, 2627-2633 (1985).

\bibitem{LRB1} A.~Lahiri, P.~K.~Roy, and B.~Bagchi,
       ''Supersymmetry in atomic physics and the radial problem,''
       J. Phys. A.: Math. Gen. {\bf 20}, 3825-3832 (1987).

\bibitem{LRB2} A.~Lahiri, P.~K.~Roy, and B.~Bagchi,
       ''Supersymmetry and the three-dimensional isotropic oscillator problem,''
       J. Phys. A.: Math. Gen. {\bf 20}, 5403-5404 (1987).

\bibitem{Kur} K.~Kurata,
       ''On the ratio of the first two eigenvalues of perturbed harmonic
       oscillators,''
       J. Math. Anal. Appl. {\bf 204}, 227-235 (1996).

\end{thebibliography}
\end{document}